\documentclass[journal,10pt]{IEEEtran}
\usepackage{cite}
\usepackage{amsmath,amssymb,amsfonts}
\usepackage{algorithmic}
\usepackage{graphicx}
\usepackage{amsthm}
\usepackage{epsfig}
\usepackage{color}
\usepackage{subcaption}
\usepackage{lettrine}
\usepackage{float}
\usepackage{lipsum}
\usepackage{bm}
\usepackage{mathtools}
\usepackage{bbm}
\usepackage{suffix}
\usepackage[T1]{fontenc}
\usepackage[colorlinks]{hyperref} 
\usepackage[nameinlink,noabbrev]{cleveref}
\usepackage{textcomp}
\usepackage{xcolor}
\usepackage{stackengine}
\usepackage[ruled,norelsize]{algorithm2e}

\usepackage{textcomp}
\def\BibTeX{{\rm B\kern-.05em{\sc i\kern-.025em b}\kern-.08em
    T\kern-.1667em\lower.7ex\hbox{E}\kern-.125emX}}

\newtheorem{lemma}{Lemma}

\newcommand{\Q}{\ensuremath{\mathbb Q}}
\newcommand{\V}{\ensuremath{\mathbb V}}
\newcommand{\W}{\ensuremath{\mathbb W}}

\newcommand{\us}{$ \mathcal{S} $}
\newcommand{\uj}{$ \mathcal{J} $}
\newcommand{\des}{$ \mathcal{D} $}
\newcommand{\eve}{$ \mathcal{E} $}

\def \treq {\stackrel{\tiny \Delta}{=}}

\DeclarePairedDelimiterX\MeijerM[3]{\lparen}{\rparen}%
{\begin{smallmatrix}#1 \\ #2\end{smallmatrix}\delimsize\vert\,#3}

\newcommand\MeijerG[8][]{%
	G^{\,#2,#3}_{#4,#5}\MeijerM[#1]{#6}{#7}{#8}}

\WithSuffix\newcommand\MeijerG*[7]{%
	G^{\,#1,#2}_{#3,#4}\MeijerM*{#5}{#6}{#7}}

\makeatletter
\def\@seccntformat#1{\@ifundefined{#1@cntformat}%
	{\csname the#1\endcsname\quad}
	{\csname #1@cntformat\endcsname}
	}
\makeatother

\begin{document}

\title{\hspace{-2mm}Improving PHY-Security of UAV-Enabled Transmission with Wireless Energy Harvesting:\\ Robust Trajectory Design and {Communications Resource Allocation}}
	

\author{
\IEEEauthorblockN{
Milad Tatar Mamaghani\IEEEauthorrefmark{1}, Yi Hong\IEEEauthorrefmark{1}, ~\IEEEmembership{Senior Member,~IEEE}}
		
\IEEEauthorblockA{
\IEEEauthorrefmark{1}Electrical and Computer Systems Engineering Department, Monash University, Melbourne, Australia}

	


\thanks{``This research was supported by the Australian Research Council under Grant DP160100528.''}}

\markboth{}{}
\maketitle

{\vspace{-7mm}}
\begin{abstract}
In this paper, we consider an unmanned aerial
vehicle (UAV) assisted communications system, including two cooperative UAVs, a wireless-powered ground destination node leveraging simultaneous wireless information and power transfer (SWIPT) technique, and a terrestrial
passive eavesdropper. One UAV delivers confidential information
to destination and the other sends jamming signals to against eavesdropping and assist destination with energy harvesting. Assuming UAVs have partial information about eavesdropper's location, we propose two transmission schemes: friendly UAV jamming (FUJ) and Gaussian jamming transmission (GJT) for the cases when jamming signals are known and unknown a priori at destination, respectively. Then, we formulate an average secrecy rate maximization problem to jointly optimize the transmission power and trajectory of UAVs, and the power splitting ratio of destination. Being non-convex and hence difficult to solve the formulated problem, we propose a computationally efficient iterative algorithm based on block coordinate descent and successive convex approximation to obtain a suboptimal solution. Finally, numerical results are provided to substantiate
the effectiveness of our proposed multiple-UAV schemes, compared to other existing
benchmarks. Specifically, we find that the FUJ demonstrates significant secrecy performance improvement in terms of the optimal instantaneous and average secrecy rate compared to the GJT  and the conventional single-UAV counterpart.
\end{abstract}


\IEEEpeerreviewmaketitle

\section{Introduction}
\lettrine[lines=2]{R}{ecently}, unmanned aerial vehicle (UAV) has been deemed as a promising wireless service provider alongside with plethora of other civilian applications (see \cite{Hayat2016sur, Zeng2016Wir, Mozaffari2019Tutorial, Li2018Sur} and references therein). This is driven by  advances in wireless equipment miniaturization as well as the economic ease of deployment and flexibility of UAVs inasmuch as various Tech giants (e.g. Facebook and Google) \cite{facebook2014} have been focusing on establishing massive UAV-assisted networks for ubiquitous connectivity. As a matter of fact, the upsurge of UAV applications in wireless communications is double-edged sword; in that bringing new opportunities and facilitating novel technologies, while accompanying with undeniable critical challenges when employed in the real world.  

On the one hand, with an increasing demand of Internet-of-things (IoT) applications, {UAVs equipped with various types of sensors, cameras, GPS, and so on, can be regarded as good candidates to serve as aerial base stations/legitimate terminals/mobile relaying and even power beacons for prolonging energy-constraint IoT devices \cite{Wang2019UAV, Perera2018Sim,Wang2018aut,Long2019Sur,sun2019Swipt}}.
In such applications, a challenging issue is that how to prolong device lifetime due to limited access to power resources and/or infrequent battery replacements \cite{Perera2018Sim}. To tackle this problem, apart from conventional energy harvesting techniques, 
simultaneous wireless information and power transfer (SWIPT) has recently emerged \cite{grover2010shannon}. To be specific, 
SWIPT captures both data and energy from the same radio frequency (RF) signal and converts into direct current for battery recharging, which enables energy harvesting in a controllable manner. This characteristic is particularly important for UAV applications to guarantee replenishable-energy ground nodes considering their dynamic adjustment capability  \cite{sun2019Swipt, Wang2018aut, Sun2019PLS, Zhaohui2020Energy}. {Specifically, a SWIPT-based UAV-aided relaying scenario
to transmit power and confidential information to an energy-constrained ground user  has been analyzed in terms of average achievable secrecy rate and energy coverage probability in \cite{Sun2019PLS}, while the secrecy rate lower bound optimization problem of such setup has been conducted in \cite{sun2019Swipt}. Aiming at minimization of the UAV's total power consumption, the authors in \cite{Zhaohui2020Energy} also explored a non-security  UAV-based wireless communications system with energy harvesting to enable data transmission of ground users in both half duplex and full duplex modes using the harvested energy. }

On the other hand, safeguarding such wireless communications system is of the most paramount challenges due to the broadcast nature of transmission and mobility of UAVs. To guarantee security of UAV communications, physical-layer (PHY) security \cite{Hayat2016sur, Yener2015PLS, Xianggong2019Resource, Sun2019Physical, Li2020PLS, Sun2019Power, Zhang2017sec, Li2018Uav, Lee2018UAV, Zhou0218Imp} have, providentially, been ascertained as a promising and computationally-efficient information secrecy approach. {For example, the resource allocation problem for a UAV-assisted secure SWIPT system is investigated in \cite{Xianggong2019Resource}. The authors in \cite{Sun2019Power} also considered the PLS of a four-node setup with UAV-enabled relaying where the eavesdroppers are distributed in a certain area with partially known location information and then studied the power allocation problem of the source and relay. } Amongst various PHY-security techniques, cooperative jamming is one viable anti-eavesdropping strategy via collaboratively transmitting jamming signals to degrade  wiretap channel quality. In \cite{Zhang2017sec}, the authors have considered a mobile UAV serving as a flying base station delivering data to a ground node in the presence of a passive eavesdropper. In \cite{Li2018Uav}, leveraging the mobility of a UAV, the authors have studied the achievable secrecy rate via trajectory design and power control optimization, and showed its improvement over conventional static jammers. This is due to the fact that the mobility of UAV-jammer allows an opportunistically jamming at a closer distance to  the eavesdropper. In \cite{Lee2018UAV}, the authors have tackled maximizing the minimum secrecy rate of jammer-incorporated UAV communications via a joint optimization of trajectory and transmit power of UAVs. In \cite{Zhou0218Imp}, the authors have studied the problem when a UAV is employed as friendly jammer to assist secure communication in the presence of unknown eavesdropper location, and they have examined the UAV-jammer displacement and power control to guarantee good reliability and security.

Motivated by above research, in this paper, we consider two flying cooperative UAVs as well as a ground destination node equipped with wireless RF energy harvester, in the presence of a passive ground eavesdropper. One UAV acts as source transmitting confidential information to destination while the other UAV broadcasts jamming signals to assist anti-eavesdropping and energy harvesting of the destination node. Note that, different from \cite{Li2019Coo, Zhang2017sec,  Li2018Mob}, we here consider a {\em SWIPT-enabled receiver} at destination for security and energy scavenging. Also, different from \cite{Zhong2018sec, Li2018Mob} wherein UAVs know the exact location of eavesdropper, we here assume that UAVs have only {\em partial information} of eavesdropper's location.
Following our setting, we make the following contributions in the paper.
\begin{itemize}
    \item We propose two cooperative UAV-jamming PHY-security schemes: {\em friendly UAV jamming} (FUJ) and {\em Gaussian jamming transmission} (GJT). In particular, in FUJ, UAV transmits jamming signals that are known a priori at destination, while in GJT, destination node has no prior information of the jamming signals from UAV.
    \item Via trajectory discretization
approach, we formulate an average secrecy rate
(ASR) optimization problem, which is challenging to solve due to non-smooth and non-concave objective function and non-convex feasible set. 
\item To make the optimization problem tractable, we propose an efficient iterative algorithm based on block coordinate descent (BCD) and successive convex approximation (SCA) methods in order to find a unique sub-optimal solution to the problem with guaranteed convergence.
\item Via the proposed iterative algorithm, we conduct optimization of the following sub-problems: transmit power of UAVs, power splitting ratio in SWIPT, as well as UAVs trajectory. 
\item We compare by simulations secrecy and energy harvesting performance, 
transmit power of UAVs of our proposed schemes under various  scenarios, demonstrating its significant performance
improvement over conventional without-jamming (WoJ) scheme, wherein there exists no UAV-jammer.
\end{itemize}

The rest of the paper is organized as follows. Section~\ref{Sec:SysMod}
introduces system model. Section~\ref{Sec:PropPLS} presents two
2-UAV transmission schemes via  cooperative UAV jamming. In Section~\ref{Sec:ProFormulation}, we formulate ASR optimization problem via trajectory discretization
approach and provide solutions in Section~\ref{Sec:ProSolu}. Simulation results are given in Section~\ref{Sec:Simu}, followed by conclusions in Section~\ref{Sec: Conl}.

\section{System Model} \label{Sec:SysMod}
\begin{figure}[t]
	\center{\includegraphics[ width= 0.85\columnwidth ]{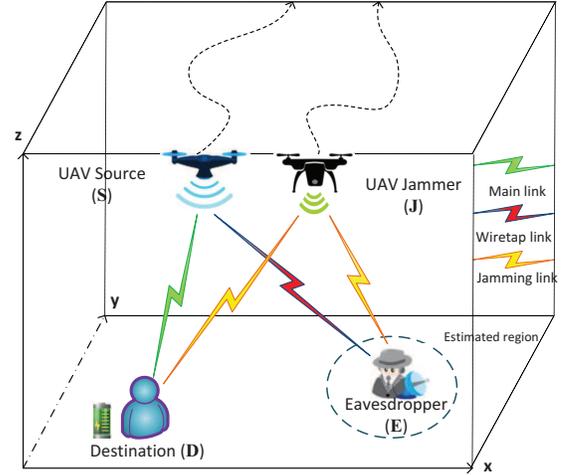}}
	\caption{\label{fig0} System model of UAV-enabled secure information and power transfer. }                                         
\end{figure}

We consider a UAV-enabled wireless communications system (see Fig. \ref{fig0}), where a UAV-source (\us) flies from initial to final locations to deliver confidential information to a legitimate {\em ground destination} (\des) in the presence of a {\em ground eavesdropper} (\eve) with {\em unknown location}. Here, we consider~\des~to be an energy-limited IoT device that is capable of harvesting energy from ambient radio resources and its receiver adopts power splitting architecture for simultaneous energy scavenging and data decoding with a power splitting ratio (PSR) $\zeta$ ($0 \leq \zeta \leq 1$)  \cite{Perera2018Sim, Milad2019UAVAccess}. Finally, a {\em UAV-jammer} (\uj) is employed to transmit {\em noise-like} jamming signals cooperatively to improve security and power the energy-constraint~\des. 

We consider that all nodes have {\em single omnidirectional antenna} that operate in {\em half-duplex mode}. We define {\em main link} (\us-\des), {\em wiretap link} (\us-\eve), {\em jamming link} (\uj-\des,~\uj-\eve), as shown in Fig. \ref{fig0}.





\subsection{System Parameters}\label{Subsec:Syspara}
Without loss of generality, we assume that all the nodes are located in a three-dimensional Cartesian coordinate system with the following parameters:
\begin{itemize}
    \item \des~has the {\em horizontal coordinate} $\W_D\in\mathbb{R}^{2\times1}$ with zero altitude,
    \item \us~and~\uj's initial and final locations corresponding to the prespecified launching and landing sites of the UAVs are  $\Q_{SI}\in\mathbb{R}^{2\times1}$, $\Q_{JI}\in\mathbb{R}^{2\times1}$, $\Q_{SF}\in\mathbb{R}^{2\times1}$, and $\Q_{JF}\in\mathbb{R}^{2\times1}$, with {\em constant flying altitude}\footnote{Indeed, one justification from a practical viewpoint behind this constant UAVs' flying altitude assumption is to guarantee the safety consideration like collision avoidance with buildings or terrain, and also more importantly, for energy consumption reduction when ascending or descending of UAVs, e.g.,  \cite{Li2019Coo}, \cite{Hongliang2018Sec}.} $H$.
    \item \us~and~\uj~have the same mission time $T$, and their horizontal location at time instant $t\in[0,T]$ are $\Q_S(t) \in \mathbb{R}^{2\times1}$ and $\Q_J(t) \in \mathbb{R}^{2\times1}$,  
    \item \us~and~\uj~have a {\em safety distance} $\tilde{D}$ to avoid collision\footnote{This is different from the traditional approach in \cite{Lee2018UAV}, where different flying altitudes are allocated to each UAV to avoid possible collision.},
    \item \us~and~\uj~have total transmission power $P_S^{tot}$ and jamming power $P_J^{tot}$, respectively, whereas at $t\in[0,T]$, their associate instantaneous powers are $P_S(t)$ and $P_J(t)$,
    \item PSR, denoted by $\zeta \in (0,1)$, is the {\em fraction of received power for information processing}, while (1-$\zeta$) is the {\em fraction of which to be harvested and stored for future use}. The {\em instantaneous PSR} is, therefore, denoted by $\zeta(t)$.
\end{itemize}
    

Further, we have the following assumptions on~\des~and~\eve's locations:
\begin{itemize}
    \item \des's location is known to both UAVs (e.g.~ \cite{Mukherjee2012det},\cite{Zhang2017sec}),
    \item \eve's location ($\W_E\in\mathbb{R}^{2\times1}$) is unknown, but both UAVs can approximately estimate it \cite{Cui2018Rub} in a collaborative manner. As such, we assume that~\eve's circular estimated region centered at $\hat{\W}_E \in\mathbb{R}^{2\times1}$ (namely {\em most-likely location of~\eve}) with radius $R_E\geq \| \W_E -\hat{\W}_E\|$ (namely {\em maximum estimation error}) are known to the UAVs, where $\|\cdot\|$ represents the $\mathrm{L_2}$-norm (Euclidean norm).
\end{itemize} 
\vspace{5mm}
{\textit{Remark:}} Note that according to \cite{Sun2019Physical}, the availability of the eavesdroppers’ location information can be classified into three cases: I) full position information, II) partial position information, and III) absence of position information. Case I becomes possible, when the eavesdroppers stay stationary, and UAV is equipped with an optical camera or a synthetic aperture radar to detect the eavesdropper's location, or it might be the case when the ground nodes are part of the same network with different roles; e.g., unscheduled users to receive particular information compared to the intended ones. Other method is presented in \cite{Mukherjee2012det} to obtain eavesdropper's location information  from the local oscillator power which is inadvertently leaked from its RF front-end, given coherent detection is used.  Case II occurs when the above detection is non-accurate, or when eavesdroppers have moved slightly so that the camera/aperture radar in UAV cannot perfectly obtain their location information. Case III occurs when all the above detection methods are failed and eavesdroppers hide themselves physically very well. In this paper, we have considered Case II.



\subsection{Channel Model}
Motivated by literature (see \cite{Zhong2018sec, Hongliang2018Sec},\cite{Yong2019Acc, Zhou2018uav, Zeng2016thr, Zeng2018uav}), in this work, we adopt a {\em probabilistic line-of-sight} (LOS) channel model that models both LoS and Non-LoS  propagations by taking
into account their occurrence probabilities \cite{Yong2019Acc}. Averaging over surrounding environment and small-scale fading, the expected channel power of UAV-ground (UG) links at time instant $t$ is \cite{Yong2019Acc}
\begin{align}\label{ch_model}
\hat{h}_{ag}(t)  = \hat{\beta}(\theta_{ag}(t)) d_{ag}(t)^{-\alpha},
\end{align}
with the regularized attenuation factor given by
\begin{align} \label{beta}
\hat{\beta}(\theta_{ag}(t)) \triangleq \beta_0 \left[ P_{\mathrm{LoS}}(\theta_{ag}(t)) + \kappa (1- P_{\mathrm{LoS}}(\theta_{ag}(t)))\right],  
\end{align}
where $d_{ag}(t)=\sqrt{\|\Q_a(t)-\W_g\|^2+H^2}$ represents the time varying distance between the aerial node $a$ and the ground node $g$. Moreover,  $\theta_{ag}(t) = \mathbf{tan^{-1}}\left(\frac{H}{d_{ag}(t)}\right)$ denotes the time-varing elevation-angle between those two, wherein  $a\in$\{\us, \uj\} and $g\in$\{\des, \eve\},  $\alpha$ denotes the path-loss exponent ($2 \leq \alpha \leq 4$) \cite{Goldsmith}, $\beta_0$ is the path loss at reference distance $d_0$ meter for omnidirectional antennas under LoS, i.e.,
\[
\beta_0 \triangleq 20 \log_{10}\left(\frac{\mathrm{C}}{4\pi d_0f_c}\right),~~~(\text{in~dB})
\]
where $\mathrm{C} = 3\times10^8~\mathrm{m/s}$ is the speed of light and $f_c$ is the carrier frequency \cite{Goldsmith}. The parameter $\kappa$ is the additional attenuation factor characterizing Non-LoS propagation (in practice it is a random variable with log-normal distribution denoting the shadowing effect);  however, in \eqref{beta}, this parameter is regarded to be constant following homogeneous assumption for Non-LoS environment. Here, In consistent with \cite{Zeng2019Ene, Yong2019Acc}, we assume that for the area of interest the elevation angle dependent probabilistic LoS function \[P_{\mathrm{LoS}}(\theta(t)) = \frac{1}{1+k_1\exp(-k_2(\theta(t)-k_1))},\]
with environmental constants $k_1, k_2>0$ follow homogeneity, leading ultimately to $\hat{\beta}(\theta(t)) \approx \bar{\beta}$ for the sake of simplicity of trajectory and resource allocation design\footnote{It is worth pointing out that this approximated and simplified model is too fruitful in some applications such as post-disaster area  wherein it is non-trivial to categorize the environment based on which the probabilistic model has been developed. However, the minimum and maximum values of path-loss component $\alpha$ can be used for upper and lower bound performance \cite{Zeng2016Wir}.}.


\section{Proposed PHY-security Schemes and Instantaneous/Average Secrecy Capacity} \label{Sec:PropPLS}
In this work, we present two PHY-security schemes involving two UAVs. Major difference between our schemes and other known two-UAV schemes (e.g. \cite{Li2018Mob, Zhong2018sec, Lee2018UAV}) lies in that {\em the additional cooperative UAV conducts not only jamming transmission but also powering~\des~in a more practical channel modelling}: 
\begin{itemize}
\item A FUJ scheme, wherein FUJ transmits jamming signals that are known a priori at~\des~ 
\item A GJT scheme, wherein~\des~has no prior knowledge\footnote{Note that while the FUJ scheme requires a priori to generate jamming signals at~\uj~and also costs a higher computational complexity at~\des~to operate jamming cancellation, it can be implemented via various approaches such as key-assisted coding; i.e., an intelligent combination of conventional cryptography with PHY-security \cite{Harrison2011cod}. Specially, when the location of the eavesdropper~\eve~is unknown to the legitimate nodes and the wiretap link quality might experience a better channel condition compared to the main link, the former scheme is capable of PHY-security enhancement, while the latter lacks such an undeniable performance advantage nonetheless provides a low complex implementation approach.} of the noise-like jamming signal.
\end{itemize}


To evaluate performance of above schemes (particularly in later simulations), we consider a {\em benchmark scheme}:
\begin{itemize}
    \item {\em No additional UAV-jammer} (WoJ) scheme with SWIPT at destination. Note that this setup is similar to \cite{Zhang2017sec}, except  \cite{Zhang2017sec} has no SWIPT.
\end{itemize}

\subsection{Instantaneous Secrecy Rate (ISR)}

 Recall system parameters in subsection \ref{Subsec:Syspara} and assume normalized bandwidth in all links. 
 
GJT has the achievable average rate over the random channel realizations at time instant $t$ as
\begin{align}\label{IMU}
\hspace{-2mm}{I}_M(t)\hspace{-1mm} =\hspace{-1mm}\log_2\hspace{-1mm}\left(\hspace{-1mm}1\hspace{-1mm}+\hspace{-1mm}\frac{\gamma_S(t) \zeta(t) \left(\|\Q_S(t)  \hspace{-1mm}- \hspace{-1mm} \W_D \|^2 \hspace{-1mm}+\hspace{-1mm} H^2\right)^{-\frac{\alpha}{2}}}{\gamma_J(t) \zeta(t) \left(\|\Q_J(t)  \hspace{-1mm}- \hspace{-1mm} \W_D \|^2 \hspace{-1mm}+\hspace{-1mm} H^2\right)^{-\frac{\alpha}{2}} \hspace{-1mm}+\hspace{-1mm} 1}\hspace{-1mm}\right),
\end{align}
where $\gamma_S(t) \treq \frac{P_S(t) \bar{\beta}}{N_0}$ and $\gamma_J(t) \treq \frac{P_J(t) \bar{\beta}}{N_0}$ with $N_0$ being the noise power at the receiver of~\des. Since the UAV jamming signal is known a priori by~\des~as well as the channel state information (CSI) is available, it can be removed from the received signals. Therefore, FUJ has the achievable instantaneous ensumble rate from~\us~to~\des~as 
\begin{align}\label{IMF}
{I}_M(t) \hspace{-1mm}=\hspace{-1mm} \log_2\left(1\hspace{-1mm}+\hspace{-1mm}\gamma_S(t) \zeta(t) \left(\|\Q_S(t) - \W_D \|^2 \hspace{-1mm}+\hspace{-1mm} H^2\right)^{-\frac{\alpha}{2}}\hspace{-1mm}\right).
\end{align}

Additionally, for both GJT and FUJ, the exact instantaneous wiretap channel capacity $\hat{I}_E(t)$  at eavesdropper can be obtained as
\begin{align}\label{IE}
\hat{I}_E(t) \hspace{-1mm}=\hspace{-1mm} \log_2\hspace{-1mm}\left(\hspace{-1mm}1\hspace{-1mm}+\hspace{-1mm}\frac{\gamma_S(t) \left(\|\Q_S(t) - \W_E \|^2 + H^2\right)^{-\frac{\alpha}{2}}}{\gamma_J(t) \left(\|\Q_J(t) - \W_E \|^2 + H^2\right)^{-\frac{\alpha}{2}} \hspace{-1mm}+\hspace{-1mm} 1}\hspace{-1mm}\right),
\end{align}
where the AWGN noise power at~\eve~is considered identical to that at~\des~for the simplicity of exposition. 

{The maximum achievable data rate by~\eve, denoted by ${I}^{max}_{E}(t)$, within the uncertainty region $R_E$, which serves as an upper-bound for the case of exact location of~\eve, can be calculated, by considering the worst-case estimation scenario by two UAVs, as
\begin{align}\label{IE_max}
{I}^{max}_{E}(t)\hspace{-1mm}=\hspace{-1mm} \log_2\hspace{-1mm}\left(\hspace{-1mm}1\hspace{-1mm}+\hspace{-1mm}\frac{\gamma_S(t)\hspace{-1mm} \left(\hspace{-1mm}\left(\|\Q_S(t)\hspace{-1mm}-\hspace{-1mm}\hat{\W}_E \|\hspace{-1mm}-\hspace{-1mm}R_E\right)^2\hspace{-2mm}+\hspace{-1mm}H^2\right)^{-\frac{\alpha}{2}}}{\gamma_J(t)\hspace{-1mm}\left(\hspace{-1mm}\left(\|\Q_J(t) \hspace{-1mm}-\hspace{-1mm} \hat{\W}_E \| \hspace{-1mm}+\hspace{-1mm} R_E\right)^2 \hspace{-2mm}+\hspace{-1mm} H^2\right)^{-\frac{\alpha}{2}} \hspace{-3mm}+\hspace{-1mm} 1}\hspace{-1mm}\right).  
\end{align}}
\begin{proof}
Please see Appendix A. 
\end{proof}

\subsection{Average Secrecy Capacity}
The achievable ASR from~\us~to~\des~with normalized transmission bandwidth is defined in bits/s/Hz as \cite{Gopala2008On}
\begin{align}\label{rsec}
\bar{R}_{sec} = \frac{1}{T}\int_{0}^{T} \left[I_M(t) - {I}^{max}_{E}(t)\right]_+\,dt,
\end{align}
where $[x]_+ = \max\{x,0\}$ and $I_M(t)$ for GJT and FUJ schemes are given in \eqref{IMU} and \eqref{IMF}, respectively, ${I}^{max}_{E}(t)$ is in \eqref{IE_max} for cooperative jamming. Note that ${I}^{max}_{E}(t)$ for WoJ is identical to \eqref{IE_max} but with setting $\gamma_J(t)=0$.

\section{Problem Formulation for Maximizing ASR}\label{Sec:ProFormulation}
To maximize \eqref{rsec}, we need a joint design of UAV trajectory, transmission power allocations, and power splitting ratio. To make our design practically feasible, we consider the {\em trajectory discretization} approach dividing the mission time $T$ into $N$ equally-spaced time slots \[
{\delta_t} \treq \frac{T}{N},\] Given ${\delta_t}$, assuming  distance variation between any UAV and the ground terminals is adequately small, we adopt {\em constant average channel gains} per slot. Other system design parameters and definitions are quantized accordingly and being constant within each time slot. Hence, our problem of interest with variables  $\mathbf{P_S} \treq \{{P}_S[n]\}^N_{n=1}$, ${\mathbf{P_J}}\treq\{P_J[n]\}^N_{n=1}$, $\pmb{\zeta}\treq\{\zeta[n]\}^N_{n=1}$, $\mathbf{Q_S} \treq \{\Q_S[n]\}^N_{n=1}$, and  $ \mathbf{Q_J} \treq \{\Q_J[n]\}^N_{n=1}$ is formulated as 
\begin{align}\label{opt_problem}
{{{\bar{R}}_{sec}^{opt}\left(\mathbf{P}^\star_\mathbf{S},\mathbf{P}^\star_\mathbf{J}, \pmb{\zeta}^\star, \mathbf{Q}^\star_\mathbf{S}, \mathbf{Q}^\star_\mathbf{J}\right)}}
=\stackrel{}{\underset{}{~\mathrm{maximize}~\frac{1}{N} \sum_{n=1}^{N}\hspace{-1mm}\left[{\tilde{R}}_{sec}[n]\right]_+ }}\nonumber\\
&\hspace{-35mm}\text{s.t.}~~~\mathrm{C1-C15},
\end{align}

with $\tilde{R}_{sec}[n]$ given by \eqref{objfunc} (shown on top of the next page)
\begin{figure*}[t]
\begin{align}\label{objfunc}
\tilde{R}_{sec}[n] &= \log_2\left(1+\frac{\zeta[n]P_S[n]\bar{\beta} \left(\|\Q_S[n] -\W_D\|^2+H^2\right)^{-\frac{\alpha}{2}}}{\zeta[n]P_J[n] \bar{\beta}\left(\|\Q_J[n] -\W_D\|^2+H^2\right)^{-\frac{\alpha}{2}}+N_0}\right)\nonumber\\
&-\log_2\left(1+\frac{P_S[n]\bar{\beta} \left(\|\Q_S[n] - \hat{\W}_E \|^2-2R_E \|\Q_S[n] - \hat{\W}_E \|+ \tilde{H}^2\right)^{-\frac{\alpha}{2}}}{P_J[n]\bar{\beta} \left(\|\Q_J[n]- \hat{\W}_E \|^2 +2R_E \|\Q_J[n] - \hat{\W}_E \|+ \tilde{H}^2\right)^{-\frac{\alpha}{2}} +N_0}\right),
\end{align}
\noindent\rule{\textwidth}{.5pt}
\end{figure*}
wherein $\tilde{H}^2 \treq R_E^2+H^2$. The constraints C1-C4 are
\begin{align}
 \mathrm{C1}:~&\frac{1}{N} \sum_{n=1}^{N} P_S[n] \leq \bar{P}_S, ~~\mathrm{C2}:~0 \leq P_S[n] \leq \hat{P}_S, \\
\mathrm{C3}:~&\frac{1}{N} \sum_{n=1}^{N} P_J[n] \leq \bar{P}_J, ~~ \mathrm{C4}:~0 \leq P_J[n] \leq \hat{P}_J,
\end{align}
where (C1,C3) and (C2,C4) are constraints of average and maximum transmission/jamming powers per time slot at~\us~and~\uj, i.e., ($\bar{P}_S$,~$\bar{P}_J$) and ($\hat{P}_S$, $\hat{P}_J$), respectively, where \[
\bar{P}_S \treq P^{tot}_S/N,~~~\bar{P}_J \treq P^{tot}_J/N.\]
Additionally, these fixed powers are chosen subject to {\em the peak to average power ratio} (PAPR) constraint, i.e., $\frac{\hat{P}_K}{\bar{P}_K}$ is restricted due to hardware limitations, where $K\in\{S,J\}$, and {\em maximum network  transmission power} per time slot as 
\begin{equation}\label{Eq:MaxTxPowerofall}
    \hat{P}_R = \hat{P}_S+\hat{P}_J.
\end{equation}
To ensure a sufficient discretization as well as valid assumptions of invariant channel condition and unchanged distance between any UAV and ground nodes, we have mobility constraints as
\begin{align}
\mathrm{C5}:~&{\Q}_S[1] = {\Q}_{SI}, \nonumber\\
\mathrm{C6}:~&\|{\Q}_S[n+1] - {\Q}_S[n]\| \treq \V_S[n]\delta_t \leq \tilde{d}_\delta,~n=1\cdots N\hspace{-1mm}-\hspace{-1mm}1 \nonumber\\
\mathrm{C7}:~&\|{\Q}_{SF} \hspace{-1mm}-\hspace{-1mm} {\Q}_S[N]\| \treq \V_S[N]\delta_t \leq \tilde{d}_\delta, 
\end{align}
and 
\begin{align}
\mathrm{C8}:~&{\Q}_J[1] = {\Q}_{JI}, \nonumber\\
\mathrm{C9}:~&\|{\Q}_J[n+1] - {\Q}_J[n]\| \treq \V_J[n]\delta_t \leq \tilde{d}_\delta,~n=1\cdots N\hspace{-1mm}-\hspace{-1mm}1 \nonumber\\
\mathrm{C10}:~&\|{\Q}_{JF} - {\Q}_J[N]\| \treq \V_J[N]\delta_t \leq \tilde{d}_\delta,
\end{align}
where $\V_S[n]$ and $\V_J[n]$ are constant speeds of~\us~and~\uj~in time slot $n$, but the velocities may vary from one slot to next. In particular, the {\em maximum horizontal displacement} of~\us~and~\uj~per slot is bounded by {\em  threshold maximum distance} $\tilde{d}_\delta \ll H$. 
For the considered two-UAV system, collision avoidance is represented by
\begin{align}
\mathrm{C11}:~~&\|{\Q}_S[n] - {\Q}_J[n]\| \geq \tilde{D},  
\end{align}
where $\tilde{D}$ is the safety distance between the two UAVs. Then, the permitted flying zone for UAVs is assumed to be a circular region with radius $\tilde{R}$, i.e.,
\begin{align}
\mathrm{C12}:~~&\|{\Q}_S[n]-\W_D\|\leq \tilde{R},\nonumber\\
\mathrm{C13}:~~&\|{\Q}_J[n]-\W_D\|\leq \tilde{R},
\end{align}
where 
\begin{align}\label{feasibility}
\tilde{R} \leq \sqrt{\left(\frac{\hat{P}_R\bar{\beta}}{\Psi_H}\right)^{\frac{2}{\alpha}} -H^2},
\end{align}
must be satisfied to avoid power outage and guarantee the viability of energy harvesting. In (\ref{feasibility}), $\Psi_H$ is the {\em minimum required input power for energy harvesting}, and $\hat{P}_R$ is given in (\ref{Eq:MaxTxPowerofall}). Finally, energy harvesting constraints are 
\begin{eqnarray}\label{Eq:EnerHarv}
&&\mathrm{C14}:~~ 0 \leq \zeta[n] < 1,\nonumber\\
  &&\mathrm{C15}:~~ \tilde{E}_H[n]\geq \Psi_H,~~\forall n
\end{eqnarray}
with harvested power in time slot $n$ given by \eqref{eh} (see top of the next page)
\begin{figure*}
\begin{align}\label{eh}
\tilde{E}_H[n] &\treq \eta  \left(1-\zeta[n]\right) \Big[
 P_S[n] \bar{\beta}\left(\|\Q_S[n] -\W_D\|^2+H^2\right)^{-\frac{\alpha}{2}}
+ P_J[n] \bar{\beta} \left(\|\Q_J[n] -\W_D\|^2+H^2\right)^{-\frac{\alpha}{2}} + N_0\Big],
\end{align}
\noindent\rule{\textwidth}{.5pt}
\end{figure*}
where $\eta$ is {\em power conversion efficiency factor}, $\zeta[n]$ represents the discretized PSR for information processing at~\des, and   $(1-\zeta[n])$ for energy harvesting.


\section{Problem solution to Maximize ASR} \label{Sec:ProSolu}
Note that \eqref{opt_problem} is non-convex and challenging to solve due to non-convex objective function, non-smooth operator, $[\cdot]_+$, and some non-convex constraints. However, at the optimal point, $\tilde{R}_{sec}[n]$ in \eqref{objfunc} should be non-negative; otherwise, by setting $P_S[n]=0$ yields $\tilde{R}_{sec}[n]=0$ ({It should be pointed out that due to inequality $\max\{x,0\} \geq x$, the resultant smooth objective function given by \eqref{objfunc} always serves a lower-bound for the objective function of the problem \eqref{opt_problem}}). Thus, our optimization problem can be turned into a non-convex yet smooth (differentiable) problem as
{
\begin{align}\label{P1}
(P1):& \stackrel{}{\underset{\mathbf{P_S,~P_J},~\pmb{\zeta},~\mathbf{\Q_S, ~\Q_J}}{\mathrm{maximize}} ~ \frac{1}{N} \sum_{n=1}^{N} {\tilde{R}}_{sec}[n] } \nonumber\\
&~~~~~~\text{s.t.}~~~~~ \mathrm{C1-C15}.
\end{align}}
The facts that, $(P1)$ is non-convex and the optimization parameters are tightly coupled due to C15, make the problem intractable and motivates us to propose an alternating optimization approach:  
an efficient iterative algorithm based on {\em block coordinate descent} (BCD) and {\em successive convex approximation} (SCA) methods, where at each iteration a single block of variables is optimized by convex optimization approach, while the remaining variables remain unchanged. By doing so, the convergence of the proposed approach to at least a {\em sub-optimal solution} is guaranteed under a feasible set \cite{Meisam2012BCD}. The remaining analysis are given as follows.
\subsection{Optimal Transmit Power of UAV-source}\label{srcoptpow}
In the following, we optimize the power allocation of~\us~for GJT, FUJ, and WoJ, under the given feasible trajectories and PSRs. Thus, the sub-problem for optimal transmission of~\us~for the most general case (GJT) can be obtained by reformulating $(P1)$ equivalently as
\begin{align}\label{ospa}
\hspace{-1mm}(P2):&\stackrel{}{\underset{\mathbf{P_S}}{\mathrm{maximize}}\sum_{n=1}^{N}~\hspace{-1mm}{\left[\log\left({1\hspace{-1mm}+\hspace{-1mm}A_nP_S[n]}\right) \hspace{-1mm}-\hspace{-1mm} \log\left({1\hspace{-1mm}+\hspace{-1mm}B_nP_S[n]}\right) \right]}} \nonumber\\
&~~~~~\text{s.t.}~~~~~ \mathrm{C1}~~\text{and}~~\mathrm{C2},\nonumber\\
&~~~~~~~~~~~~~\widetilde{\mathrm{C15}}:~C_nP_S[n]+D_n \geq \Psi_H,~\forall n
\end{align}
where $\log(\cdot)$ represents natural logarithm, the auxiliary constants  $\{A_n\}_{n=1}^N$, $\{B_n\}_{n=1}^N$, $\{C_n\}_{n=1}^N$, and $\{D_n\}_{n=1}^N$, are given by \begin{align}\label{ak_ujt}
A_n  = \frac{{\gamma_0\zeta[n] \left(\|\Q_S[n] -\W_D\|^2+H^2\right)^{-\frac{\alpha}{2}}}}{{\zeta[n]\gamma_J[n] \left(\|\Q_J[n] -\W_D\|^2+H^2\right)^{-\frac{\alpha}{2}}+1}},
\end{align}
\begin{align}\label{bk_ujt}
B_n  \hspace{-1mm}=\hspace{-1mm} \frac{\gamma_0\hspace{-1mm}\left(\|\Q_S[n] \hspace{-1mm}-\hspace{-1mm} \hat{\W}_E \|^2 \hspace{-1mm}-\hspace{-1mm}2R_E \|\Q_S[n] \hspace{-1mm}-\hspace{-1mm} \hat{\W}_E \|\hspace{-1mm}+\hspace{-1mm} \tilde{H}^2\right)^{-\frac{\alpha}{2}}}{\gamma_J[n]\hspace{-1mm} \left(\|\Q_J[n] \hspace{-1mm}-\hspace{-1mm} \hat{\W}_E \|^2 \hspace{-1mm}+2R_E \|\Q_J[n] \hspace{-1mm}-\hspace{-1mm} \hat{\W}_E \|\hspace{-1mm}+\hspace{-1mm} \tilde{H}^2\right)^{-\frac{\alpha}{2}}\hspace{-4mm} +\hspace{-1mm}1},
\end{align}
\begin{align}\label{ck_ujt}
C_n = \eta \bar{\beta} \left(1-\zeta[n]\right)\left(\|\Q_S[n] -\W_D\|^2+H^2\right)^{-\frac{\alpha}{2}},
\end{align}
\begin{align}\label{dk_ujt}
\hspace{-2mm}D_n \hspace{-1mm}=\hspace{-1mm} \eta  \left(1\hspace{-1mm}-\hspace{-1mm}\zeta[n]\right) \hspace{-1mm}\left[P_J[n] \bar{\beta}  \left(\|\Q_J[n] \hspace{-1mm}-\hspace{-1mm}\W_D\|^2\hspace{-1mm}+\hspace{-1mm}H^2\right)^{-\frac{\alpha}{2}} \hspace{-2mm}+\hspace{-1mm} N_0\right],
\end{align}
where $\gamma_0 \treq\frac{\bar{\beta}}{N_0}$, $\gamma_J[n] \treq \frac{P_J[n] \bar{\beta}}{N_0}$, and $\tilde{H} = \sqrt{{R^2_E}+H^2}$. The sub-problem $(P2)$ is still non-convex with respect to $\mathbf{P_S}$ due to non-convex  objective function. Since one can readily verify that problem $(P2)$ satisfies the Slater's condition, strong duality attains which enables us to obtain the optimal solution by solving the corresponding Lagrange dual problem using Karush-Kuhn-Tucker (KKT) conditions. As such, by temporarily dropping C2 and $\widetilde{\mathrm{C15}}$, and also letting $\tilde{\mathbf{P}}_S$ and $(\tilde{\mathbf{P}}_S, \lambda)$ be any primal and dual optimal points with zero duality gap, {the Lagrangian function can be computed as
\begin{align}
\hspace{-3mm}\mathcal{L}\left(\mathbf{P_S},\lambda\right) &=\nonumber\\
&\hspace{-15mm}\sum_{n=1}^{N}\hspace{-0.5mm}\left[\log\hspace{-0.5mm}\left({1\hspace{-1mm}+\hspace{-1mm}B_nP_S[n]}\right) \hspace{-1mm}-\hspace{-1mm} \log\hspace{-0.5mm}\left({1\hspace{-1mm}+\hspace{-1mm}A_nP_S[n]}\right) \hspace{-1mm}+\hspace{-1mm}\lambda \hspace{-1mm} \left(P_S[n]\hspace{-1mm}-\hspace{-1mm}\bar{P}_S\right)\hspace{-0.5mm}\right],
\end{align}
where $\lambda \geq 0$ is the {\em Lagrange factor}. Then, maximizing the Lagrangian dual function defined as 
\[g\left(\lambda \right) \treq \inf_{\mathbf{P_S}}~\{\mathcal{L}\left(\mathbf{P_S},\lambda\right)\},\]
one can reach the optimality condition as \cite{Gopala2008On} 
\begin{align}
\frac{A_n}{{1+A_nP_S[n]}} -  \frac{B_n}{{1+B_nP_S[n]}}  - \lambda = 0,~~~\forall n
\end{align}
Solving the above equation with respect to $P_S[n]$ and also taking into account constraints C2 and $\widetilde{\mathrm{C15}}$, leads to the closed-form analytical solution for optimal UAV-source's power allocation as 
\begin{eqnarray}\label{optimalsrcpow_wj}
P^{\star}_S[n]\hspace{-1mm} =\hspace{-1mm} 
\begin{cases}
\min\hspace{-1mm}\left\{\max\hspace{-1mm}\left\{\left[\frac{\Psi_H-D_n}{C_n}\right]_+\hspace{-1mm}, \tilde{P}_S[n] \right\}\hspace{-1mm},\hat{P}_S\right\},& A_n \hspace{-1mm}\geq \hspace{-1mm}B_n\\
 \left[\frac{\Psi_H-D_n}{C_n}\right]_+,&  A_n \hspace{-1mm}<\hspace{-1mm} B_n
\end{cases}
\end{eqnarray}
where
\begin{align}\label{srcpowtilde}
\tilde{P}_S[n] \hspace{-1mm}=\hspace{-1mm} \frac{1}{2}\hspace{-1mm}\left[\sqrt{\left(\hspace{-1mm}\frac{1}{B_n}\hspace{-1mm}-\hspace{-1mm}\frac{1}{A_n}\hspace{-1mm}\right)^2 \hspace{-2mm}+\hspace{-1mm} \frac{4}{\lambda}\hspace{-1mm}\left(\hspace{-1mm}\frac{1}{B_n}\hspace{-1mm}-\hspace{-1mm}\frac{1}{A_n}\hspace{-1mm}\right) } \hspace{-1mm}-\hspace{-1mm} \left(\hspace{-1mm}\frac{1}{B_n}\hspace{-1mm}+\hspace{-1mm}\frac{1}{A_n}\hspace{-1mm}\right)\hspace{-1mm}\right],
\end{align}}
wherein the non-negative  Lagrange factor $\lambda$ can be obtained by applying a simple bisection search such that the UAV's source power budget constraint; i.e., \[\sum_{n=1}^{N}P^\star_S[n] \leq P^{tot}_S,\]
is satisfied.

We note that, for FUJ, the optimal~\us~power allocation $\mathbf{P}^{\star}_S$, following the similar approach to GJT, can be obtained as \eqref{optimalsrcpow_wj} by removing the term  $\zeta[n]\gamma_J[n] \left(\|\Q_J[n] -\W_D\|^2+H^2\right)^{-\frac{\alpha}{2}}$ from denominator of \eqref{ak_ujt}. Likewise, the optimal~\us~power allocation for the WoJ is given by \eqref{optimalsrcpow_wj} 
by letting $P_J[n]$ equals to zero in \eqref{ak_ujt}, \eqref{bk_ujt}, and \eqref{dk_ujt}.

\subsection{Optimal Transmit Power of UAV-jammer}\label{jmroptpow}
Under keeping other variables unchanged, we aim at optimizing the jamming transmit power for GJT and FUJ. As such, the sub-problem for optimization of the transmit power of~\uj~for GJT can be obtained by rewriting $(P1)$ as
\begin{align}\label{ojpa}
\hspace{-1mm}(P3):& \stackrel{}{\underset{\mathbf{P_J}}{\mathrm{maximize}}\sum_{n=1}^{N}\hspace{-0.5mm}\log\hspace{-1mm}\left(\hspace{-1mm}1\hspace{-1mm}+\hspace{-1mm}\frac{A_n}{B_nP_J[n]\hspace{-1mm}+\hspace{-1mm}1}\hspace{-1mm}\right) \hspace{-1mm}-\hspace{-1mm} \log\hspace{-1mm}\left(\hspace{-1mm}1\hspace{-1mm}+\hspace{-1mm}\frac{C_n}{D_nP_J[n]\hspace{-1mm}+\hspace{-1mm}1}\hspace{-1mm}\right)} \nonumber\\
&~~~\text{s.t.}~~~\mathrm{C3},~\mathrm{\tilde{C}4}: \left[\hspace{-1mm}\frac{\Psi_H \hspace{-1mm}-\hspace{-1mm}E_n}{F_n}\hspace{-1mm}\right]_+ \hspace{-2mm}\leq \hspace{-0.5mm}P_J[n]\hspace{-0.5mm} \leq \hspace{-0.5mm}\hat{P}_J,~\forall n
\end{align}
where the auxiliary constants $\{A_n\}_{n=1}^N$, $\{B_n\}_{n=1}^N$, $\{C_n\}_{n=1}^N$, $\{D_n\}_{n=1}^N$, $\{E_n\}_{n=1}^N$, $\{F_n\}_{n=1}^N$ are taken as
\begin{align}
\hspace{-2mm}A_n\hspace{-1mm} &=\hspace{-1mm} {{\zeta[n] \gamma_S[n] \left(\|\Q_S[n] -\W_D\|^2+H^2\right)^{-\frac{\alpha}{2}}}}, \label{ak_ujt_ojpa} \\
\hspace{-2mm}B_n\hspace{-1mm}  &=\hspace{-1mm}  \gamma_0 \zeta[n]\left(\|\Q_J[n] -\W_D\|^2+H^2\right)^{-\frac{\alpha}{2}},  \label{bk_ujt_ojpa} \\
\hspace{-2mm}C_n\hspace{-1mm} &=\hspace{-1mm} \gamma_S[n]\hspace{-1mm} \left(\|\Q_S[n] \hspace{-1mm}-\hspace{-1mm} \hat{\W}_E \|^2 \hspace{-1mm}-\hspace{-1mm}2R_E \|\Q_S[n] \hspace{-1mm}-\hspace{-1mm} \hat{\W}_E \|\hspace{-1mm}+\hspace{-1mm} \tilde{H}^2\hspace{-1mm}\right)^{-\frac{\alpha}{2}}\hspace{-4mm}, \label{ck_ujt_ojpa} \\
\hspace{-2mm}D_n\hspace{-1mm} &=\hspace{-1mm}  \gamma_0\hspace{-1mm}\left(\|\Q_J[n] \hspace{-1mm}-\hspace{-1mm} \hat{\W}_E \|^2 \hspace{-1mm}+2R_E \|\Q_J[n]\hspace{-1mm} -\hspace{-1mm} \hat{\W}_E \|\hspace{-1mm}+\hspace{-1mm} \tilde{H}^2\right)^{-\frac{\alpha}{2}}\hspace{-2mm}, \label{dk_ujt_ojpa} \\ 
\hspace{-2mm}E_n\hspace{-1mm}&=\hspace{-1mm}  \eta \left(1\hspace{-1mm}-\hspace{-1mm}\zeta[n]\right)\hspace{-1mm}\left[ P_S[n]\bar{\beta} \left(\|\Q_S[n] \hspace{-1mm}-\hspace{-1mm}\W_D\|^2\hspace{-1mm}+\hspace{-1mm}H^2\right)^{-\frac{\alpha}{2}} \hspace{-2mm}+\hspace{-1mm}N_0\hspace{-0.5mm} \right]\hspace{-1mm},  \label{ek_ujt_ojpa} \\
\hspace{-2mm}F_n \hspace{-1mm}&=\hspace{-1mm}  \eta \bar{\beta} \left(1-\zeta[n]\right) \left(\|\Q_J[n] -\W_D\|^2+H^2\right)^{-\frac{\alpha}{2}}. \label{fk_ujt_ojpa}
\end{align}
The sub-problem $(P3)$ is still non-convex\footnote{Note that compared to~\us's~optimal power allocation ($P2$), which we could solve the non-convex but differentiable problem analytically via its Lagrange dual approach, the objective function of ($P4$) is quite sophisticated inasmuch as the Lagrange method leads to a harder problem to solve analytically. Therefore, here we employ another technique.} with respect to $\mathbf{P_J}$ due to non-convex objective function being in the form of convex-minus-convex based on Lemma \ref{ccv_optjmrpow} given below.
\begin{lemma}\label{ccv_optjmrpow}
Let $\mathbf{x} \in \mathbb{R}^{N\times1}$ be a vector of variables, $\{a_n\}_{n=1}^{N}$ and $\{b_n\}_{n=1}^{N}$ be all non-negative constants. Then, the vector function defined as
\begin{align}
	f(\mathbf{x}) = \sum_{n=1}^{N}\log\left(1+\frac{a_n}{b_n x[n]+1}\right),
\end{align}
is convex. 
\begin{proof}
By calculating the gradient vector and also obtaining the Hessian matrix of $f(\mathbf{x})$ we have
\begin{align}
\nabla f(\mathrm{x}) = \left\{-\frac{a_nb_n}{(1+a_n+b_nx[n])(1+b_nx[n])}\right\}_{n=1}^{N},
\end{align}
\begin{align} \label{hess_f}
\mathcal{H}_f = \mathbf{diag}\left(\frac{a_n b_n^2 (a_n+2 b_n x[n]+2)}{(b_n x[n]+1)^2 (a_n+b_n x[n]+1)^2}\right),
\end{align}
where $\nabla (\cdot)$ and $\mathcal{H}$ represent gradient and Hessian operators, respectively. The convexity of $f(\mathbf{x})$ follows from the fact that the Hessian matrix given by \eqref{hess_f} is positive semi-definite. Since it is in a  diagonal form with all non-negative elements, which further implies that all the eigenvalues corresponding to the Hessian matrix are non-negative. This completes the proof.\end{proof}
\end{lemma}
Since the first term of the objective function to be maximized is convex, our approach is two-fold: approximating this convex term with its corresponding concave lower bound, and applying SCA in an iterative manner. {By doing so, we are able to reach an approximate solution  with guaranteed convergence.} Specifically, we replace the first convex term $(P3)$ with its first order Taylor expansion at $\{P_J^k{[n]}\}_{n=1}^{N}$, which is defined as the given transmit power of~\uj~at iteration $k$. It is worth mentioning that based on first-order condition \cite{cvx_boyd}, the first order Taylor approximation at the local point $x_0 \in \mathbb{R}^{N\times1}$ provides a global under-estimator of a convex function $f(x)$, i.e.,
\begin{align}\label{key}
f(x) \geq  f(x_0) + \nabla f(x)^T (x-x_0),
\end{align}
where $(\cdot)^\dagger$ represents transpose operator. { Thus, for any given local point at iteration $k$; i.e., $\mathbf{P}^k_\mathbf{J} = \{p^k_J[n]\}_{n=1}^{N}$, $(P3)$ turns into an approximated convex problem as
\begin{align}\label{ojpa_cvx}
({P4}):& \stackrel{}{\underset{\mathbf{P_J}}{\mathrm{maximize}}\sum_{n=1}^{N}\hat{B}_n+\hat{A}_nP_J[n]-\log\left(\hspace{-1mm}1\hspace{-1mm}+\hspace{-1mm}\frac{C_n}{D_nP_J[n]\hspace{-1mm}+\hspace{-1mm}1}\hspace{-1mm}\right)} \nonumber\\
&~~~~~\text{s.t.}~~~~~ \mathrm{C3}~\text{and}~\mathrm{\tilde{C}4},~\forall n
\end{align}
where
\begin{align}
\hat{A}_n &=- \frac{A_nB_n}{(1\hspace{-1mm}+\hspace{-1mm}A_n\hspace{-1mm}+\hspace{-1mm}B_nP_J^k[n])(1+B_nP_J^k[n])},\\
\hat{B}_n&= \log\left(1+\frac{A_n}{B_nP_J^k[n]+1}\right),
\end{align}}
Note that $({P4})$ is a convex problem for which the Slater's conditions can be readily verified, any points $\mathbf{{P}^\star_J}$ and $\left(\mathbf{{P}^\star_J},\lambda^\star \right)$ satisfying the KKT conditions are primal and dual optimal with zero duality gap, which implies that the dual optimum is attained. { Although  problem $({P4})$ can be numerically solved by any standard convex
optimization techniques such as the interior-point method \cite{cvx_boyd}, we are going to step further and apply Lagrangian method to gain more insight into structure of the sub-optimal solution and also effectively reduce the complexity of the algorithm. As such, temporarily dropping the constraint $\tilde{C}4$, the Lagrange dual function is written as
\begin{align}
g\left(\mathbf{P_J},\nu\right) &=\nonumber\\
&\hspace{-15mm}\inf_{\mathbf{P_J}}\hspace{-0.5mm}\left\{\hspace{-0.5mm}\sum_{n=1}^{N}\hspace{-0.5mm}\left[-\hat{A}_nP_J[n]\hspace{-1mm}-\hspace{-1mm}\hat{B}_n\hspace{-1mm}+\hspace{-1mm}\log\hspace{-0.5mm}\left(\hspace{-1mm}1\hspace{-1mm}+\hspace{-1mm}\frac{C_n}{D_nP_J[n]\hspace{-1mm}+\hspace{-1mm}1}\hspace{-1mm}\right) \hspace{-1mm}+\hspace{-1mm}\nu\hspace{-1mm} \left(P_J[n] \hspace{-1mm}-\hspace{-1mm} \bar{P}_J\right)\hspace{-1mm}\right]\hspace{-1mm}\right\}\hspace{-1mm},
\end{align} 
where the non-negative scalar $\nu$ is the Lagrange multiplier corresponding to $\mathrm{\tilde{C}3}$ in $(P5)$.
Then, solving $\nabla g\left(\mathbf{P_J},\nu\right) =0$ results in the optimality condition as
\begin{align}
\frac{1}{P_J[n]+\frac{1+C_n}{D_n}} - \frac{1}{P_J[n]+\frac{1}{D_n}} +\nu - \hat{A}_n = 0,
\end{align}
which can be rewritten as
\begin{align}
\hspace{-4mm}{P_J}^2[n] \hspace{-1mm}+\hspace{-1mm}\left(\frac{2+C_n}{D_n}\right)P_J[n] \hspace{-1mm}+\hspace{-1mm} \left[\frac{1+C_n}{D^2_n}\hspace{-1mm}-\hspace{-1mm}\frac{C_n}{D_n(\nu-\hat{A}_n)}\right] \hspace{-0.5mm}=\hspace{-0.5mm} 0,
\end{align}
Finally, solving the equation above while considering constraint $\mathrm{\tilde{C}4}$;  we reach the optimal solution of $\mathbf{P_J^\star} = \{{P_J^\star}[n]\}_{n=1}^{N}$ as
\begin{align}\label{Pjstar}
P_J^\star[n] &=\nonumber\\
&\hspace{-10mm} \min\hspace{-0.5mm}\left\{\hspace{-0.5mm}\max\hspace{-1mm}\left\{\hspace{-0.5mm}\left[\frac{\sqrt{\frac{4C_nD_n}{\nu-\hat{A}_n}\hspace{-1mm}+\hspace{-1mm}C^2_n} \hspace{-1mm}-\hspace{-1mm} \left(C_n\hspace{-1mm}+\hspace{-1mm}2\right)}{2D_n}\right]\hspace{-0.5mm},\hspace{-0.5mm} \left[\frac{\Psi_H \hspace{-1mm}-\hspace{-1mm} E_n}{F_n}\right]_+\hspace{-1mm}\right\}\hspace{-1mm},\hspace{-0.5mm} \hat{P}_J\hspace{-1mm}\right\}.
\end{align}
where $\nu \geq 0$ is the Lagrange multiplier at optimal point, satisfying $\sum_{n=1}^{N}P^{\star}_J[n] \leq P^{tot}_J$, which can be attained by a simple bisection search. We note that $(P4)$ is a lower-bound to $(P3)$ but with the same constraints, so the solution to $(P4)$, i.e., $\mathbf{P}^{\star}_J$, is no less than that of  $(P3)$ at the given point $\left(\hat{\mathbf{B}}_n,\hat{\mathbf{A}}_n, \mathbf{P}^k_\mathbf{J}\right)$.
Similarly, for the FUJ scheme  the optimal~\uj~power allocation $\mathbf{P}^{\star}_J$ is obtained as \eqref{Pjstar} but with setting $\hat{A}_n=0$.}

\subsection{Optimal power splitting ratio}
We aim at designing an efficient power splitter at destination~\des. For fixed $\mathbf{P_K}$ and $\mathbf{Q_K}$, where $K\in $\{\us,~\uj\}, the equivalent sub-problem for optimizing PSR $\{\zeta[n]\}_{n=1}^{N}$ of both GJT and FUJ, is recasted as
\begin{align}\label{opsr}
(P6):&\stackrel{}{\underset{\pmb{\zeta}}{\mathrm{maximize}}~~\sum_{n=1}^{N}{\left[\log\left({1+\frac{A_n \zeta[n]}{B_n\zeta[n]+1}}\right) \right]}} \nonumber\\
&~~~~~\text{s.t.}~ \mathrm{\tilde{C}14}:~~0 \leq \zeta[n] \leq \left[ 1-\frac{\Psi_H}{\eta C_n}\right]_+,~\forall n
\end{align}
where the auxiliary constants for $n\in\{1,2,\cdots, N\}$ are defined as
\begin{align}
A_n  &= \gamma_S[n] \left(\|\Q_S[n] -\W_D\|^2+H^2\right)^{-\frac{\alpha}{2}}, \label{ak_ujt_opsr} \\
B_n & = \gamma_J[n] \left(\|\Q_J[n] -\W_D\|^2+H^2\right)^{-\frac{\alpha}{2}}, \label{bk_ujt_opsr}\\
C_n &= P_S[n] \bar{\beta}\left(\|\Q_S[n] -\W_D\|^2+H^2\right)^{-\frac{\alpha}{2}} 
 \nonumber\\
&+ P_J[n] \bar{\beta}\left(\|\Q_J[n] -\W_D\|^2+H^2\right)^{-\frac{\alpha}{2}} + N_0, \label{ck_ujt_opsr}
\end{align}{\tiny }
It can be verified from Lemma \ref{zetacvx} that the problem $(P6)$ is concave and its objective function is monotonically increasing. 
\begin{lemma}\label{zetacvx}
	Let $x>0$ be a scalar variable and $a$ and $b$ be positive constants. Define $f(x)= \log \left(\frac{a x}{b x+1}+1\right)$.  Taking the first and the second derivative of $f(x)$ with respect to $x$ results in $\mathrm{D}f(x)=\frac{a}{(b x+1) (a x+b x+1)}$ and $\mathrm{D}^2f(x)= -\frac{a (2 a b x+a+2 b (b x+1))}{(b x+1)^2 (a x+b x+1)^2}$ respectively, where $\mathrm{D}$ is the differentiation operator. Since for any value of $x$ in the domain of $f$ we have $\mathrm{D}f(x) > 0$ and $\mathrm{D}^2f(x) < 0$, this illustrates that the function is strictly concave being monotonic increasing.
	Besides, we know that the log-product function or equivalently $h=\sum\log(\mathbf{x})$ where $\mathbf{x} = \{x_i\}_{i=1}^{N}$ is concave and non-increasing with respect to each argument $x_i$. Therefore, from the vector composition law \cite{cvx_boyd} one can readily conclude that $g(\mathrm{x}) = h o f(\mathbf{x}) = h(f(x_1),f(x_2), \cdots, f(x_N))$ is concave.
\end{lemma}
Therefore, the analytical solution for $\zeta^\star$ for GJT scenario can be readily obtained as
\begin{align}\label{zeta_opt}
\zeta^\star[n] = \left[ 1-\frac{\Psi_H}{\eta C[n]}\right]_+,
\end{align}
For FUJ and WoJ, replacing the constants $B_n=0$ and $C_n=P_S[n]\bar{\beta} \left(\|\Q_S[n] -\W_D\|^2+H^2\right)^{-\frac{\alpha}{2}}+N_0$ with \eqref{bk_ujt_opsr} and \eqref{ck_ujt_opsr}, respectively, one can apply similar approach in \eqref{zeta_opt} to obtain the optimal solution $\pmb{\zeta}^\star$. 
\subsection{Optimal UAV-source trajectory design}
We now aim at optimizing the approximated path of~\us~offline for the three schemes in terms of ASR under given the other variables.
~The corresponding sub-problem of~\us-trajectory design for GJT is reformulated as
\begin{align}\label{osrctrj}
\hspace{-3mm}(P7):~& \stackrel{}{\underset{\mathbf{Q_S}}{\mathrm{maximize}}~\sum_{n=1}^{N}\log \Phi_1(\Q_S[n])} \nonumber\\
&~~~~~~\text{s.t.}~~~~~ \mathrm{C5-C7,~~C11-C13},\nonumber\\
&\mathrm{\tilde{C}15}:~ C_n\left(\|\Q_S[n] \hspace{-1mm}-\hspace{-1mm}\W_D\|^2\hspace{-1mm}+\hspace{-1mm}H^2\right)^{-\frac{\alpha}{2}} \hspace{-1mm}+\hspace{-1mm}
D_n \hspace{-1mm} \geq \hspace{-1mm} \Psi_H,
\end{align}
where
\begin{align}
\hspace{-5mm}\Phi_1(\Q_S[n])&=\nonumber\\
&\hspace{-15mm}\frac{1\hspace{-1mm}+\hspace{-1mm}A_n \left(\|\Q_S[n] \hspace{-1mm}-\hspace{-1mm}\W_D\|^2\hspace{-1mm}+\hspace{-1mm}H^2\right)^{-\frac{\alpha}{2}}}{1\hspace{-1mm}+\hspace{-1mm}B_n\left(\|\Q_S[n]\hspace{-1mm} -\hspace{-1mm} \hat{\W}_E \|^2 \hspace{-1mm}-2R_E \|\Q_S[n] \hspace{-1mm}-\hspace{-1mm} \hat{\W}_E \|\hspace{-1mm}+\hspace{-1mm} \tilde{H}^2\right)^{-\frac{\alpha}{2}}} ,
 \end{align}
\begin{align}
\hspace{-5mm}A_n&\hspace{-1mm}=\hspace{-1mm} \frac{\zeta[n] \gamma_S[n]}{\zeta[n]\gamma_J[n] \left(\|\Q_J[n] -\W_D\|^2+H^2\right)^{-\frac{\alpha}{2}}+1}, \label{an_ujt}\\
\hspace{-5mm}B_n&\hspace{-1mm}=\hspace{-1mm} \frac{\gamma_S[n]}{\gamma_J[n]\hspace{-1mm} \left(\hspace{-0.5mm}\|\Q_J[n] \hspace{-1mm}-\hspace{-1mm} \hat{\W}_E \|^2 \hspace{-1mm}+\hspace{-1mm}2R_E \|\Q_J[n] \hspace{-1mm}-\hspace{-1mm} \hat{\W}_E \|\hspace{-1mm}+\hspace{-1mm} \tilde{H}^2\hspace{-1mm}\right)^{-\frac{\alpha}{2}} \hspace{-5mm}+\hspace{-1mm}1}\hspace{-1mm},\label{bn_ujt}\\
\hspace{-10mm}C_n &\hspace{-1mm}=\hspace{-1mm} \eta \bar{\beta}\left(1-\zeta[n]\right)P_S[n], \label{cn_ujt} \\
\hspace{-3mm}D_n &\hspace{-1mm}=\hspace{-1mm}\eta\left(1\hspace{-1mm}-\hspace{-1mm}\zeta[n]\right)\left[ P_J[n]  \bar{\beta} \left(\|\Q_J[n] \hspace{-1mm}-\hspace{-1mm}\W_D\|^2\hspace{-1mm}+\hspace{-1mm}H^2\right)^{-\frac{\alpha}{2}} \hspace{-2mm}+\hspace{-1mm}N_0\right],\label{dn_ujt}
\end{align}
The optimization problem $(P7)$ is non-convex due to the fact that the objective function is not concave with respect to $\Q_S[n]$ and the constraints $\mathrm{{C}11}$ and $\mathrm{\tilde{C}15}$ are not convex, therefore, it is hard to solve optimally. To simplify it, we reformulate $(P7)$ by introducing the slack variables $\mathbf{T} = \{T[1], T[2], \cdots, T[N]\}$ and $\mathbf{U} = \{U[1], U[2], \cdots, U[N]\}$ and obtain 
\begin{align}\label{osrctrj_1}
(P8):~& \stackrel{}{\underset{\mathbf{Q_S}, \mathbf{T}, \mathbf{U}}{\mathrm{maximize}}}\sum_{n=1}^{N} \log \frac{1+A_nT^{-\frac{\alpha}{2}}[n]}{1+B_nU^{-\frac{\alpha}{2}}[n] } \nonumber\\
&~~~~~~\text{s.t.}~~\mathrm{C5-C7,~C11-C13},\nonumber\\
&\hspace{-12mm}\mathrm{\tilde{C}15}:~ C_nT^{-\frac{\alpha}{2}}[n] \hspace{-1mm}+\hspace{-1mm} D_n \geq \Psi_H,\nonumber\\
&\hspace{-12mm}\mathrm{C16}:~\|\Q_S[n] \hspace{-1mm}-\hspace{-1mm}\W_D\|^2\hspace{-1mm}+\hspace{-1mm}H^2 \hspace{-1mm}-\hspace{-1mm} T[n] \leq 0,\nonumber\\
&\hspace{-12mm}\mathrm{C17}:~\|\Q_S[n] \hspace{-1mm}-\hspace{-1mm} \hat{\W}_E \|^2 \hspace{-1mm}-\hspace{-1mm}2R_E \|\Q_J[n] \hspace{-1mm}-\hspace{-1mm} \hat{\W}_E \|\hspace{-1mm}+\hspace{-1mm} \tilde{H}^2 \hspace{-1mm}-\hspace{-1mm} U[n]\hspace{-1mm} \geq \hspace{-1mm}0,
\end{align}
Note that C16 must hold with equality at the optimal point, otherwise by decreasing $T[n]$ one can increase the value of objective function without violating any constraints, similarly for C17. Then $(P7)$ and $(P8)$ are equivalent and have the same optimal points. Next, based on Lemma \ref{ccv}, we observe that the objective function of $(P8)$ is in the form of convex-minus-convex.
\begin{lemma}\label{ccv}
Let define the function $f(x; a,b) = \log(1+ax^{-b})$ with non-negative parameters $a$ and $b$. Taking the first and second derivatives of the function with respect to $x$ yields
\begin{align}
\mathrm{D}f = -\frac{a b}{x \left(a+x^b\right)},~~~~\mathrm{D^2}f =\frac{a b \left(a+(b+1) x^b\right)}{x^2 \left(a+x^b\right)^2},
\end{align}
where $f(x)$ is convex as $\mathrm{D^2}f \geq 0$. Note that we implicitly take the extended-value extension of $f(x)$, i.e., $\widetilde{f}(x)$, which is defined $\infty$ outside the domain of $f(x)$ for the latter result. Thus, the summation of convex functions results in a convex function. This completes the proof.
\end{lemma}
\begin{lemma}\label{over_estimator}
	Let $\mathbf{x}$ be a vector of variables $\{x_i\}_{i=1}^{N}$ and  $\mathbf{a} \in \mathbb{R}^{N\times1}$  be a constant vector . The function of negative norm-squared of this two vectors; $f(\mathbf{x}) = -\|\mathbf{x}-\mathbf{a}\|^2$, which obviously is a concave function with respect to the vector $\mathbf{x}$, has a convex upper-bound given by
	\begin{align}
	-\|\mathbf{x}-\mathbf{a}\|^2 \leq \|\mathbf{x_0}\|^2 - 2\left(\mathbf{x_0} - \mathbf{a}\right)^\dagger\mathbf{x} - \|\mathbf{a}\|^2,
	\end{align}
	\begin{proof}
See appendix B.
	\end{proof}
\end{lemma}
Using Lemmas \ref{ccv} and \ref{over_estimator}, we reformulate $(P8)$ in an approximated convex form by having concave objective function with convex feasible set as 
\begin{align}\label{osrctrj_2}
(P9):& \stackrel{}{\underset{\mathbf{Q_S}, \mathbf{T}, \mathbf{U}}{\mathrm{maximize}}}\sum_{n=1}^{N}
\hat{A}_n T[n] -\log\left(1\hspace{-1mm}+\hspace{-1mm}B[n]U^{-\frac{\alpha}{2}}[n] \right)\nonumber\\
&~~~\text{s.t.}~~~\mathrm{C5-C7},\nonumber\\
&\hspace{-10mm}\mathrm{\tilde{C}11}:~\tilde{D}^2 \hspace{-1mm}+\hspace{-1mm} \|{\Q}^k_S\|^2 \hspace{-1mm}-\hspace{-1mm} 2\left({\Q}^k_S \hspace{-1mm}-\hspace{-1mm} {\Q}^k_J\right)^\dagger{\Q}_S[n] \hspace{-1mm}-\hspace{-1mm} \|{\Q^k_J}\|^2 \leq  0\nonumber\\
&\hspace{-10mm}\mathrm{C12-C13},~\mathrm{\widetilde{C}15}:~~ \left[\frac{\Psi_H\hspace{-1mm}-\hspace{-1mm}D_n}{C_n}\right]_+ T^{\frac{\alpha}{2}}[n]\hspace{-1mm} \leq \hspace{-1mm}1,~\mathrm{C16},\nonumber\\
&\hspace{-10mm}\mathrm{\tilde{C}17}:~2R_E \|\Q_S[n] \hspace{-1mm}-\hspace{-1mm} \hat{\W}_E \|\hspace{-1mm}-\hspace{-1mm} 2\left({\Q}^k_S \hspace{-1mm}-\hspace{-1mm} {\hat{\W}_E}\right)^\dagger{\Q_S[n]} \hspace{-1mm}-\hspace{-1mm} U[n]  \hspace{-1mm}+\hspace{-1mm} \widetilde{\mathrm{H}} \hspace{-1mm}\leq\hspace{-1mm} 0,\nonumber\\
\end{align}
where $\widetilde{\mathrm{H}} \treq  \|{\Q^k_S[n]}\|^2 - \|{\hat{\W}_E}\|^2 - \tilde{H}^2$. Besides,
\begin{align}
\hat{A}_n = -\frac{\alpha A_n}{2T_n \left(A_n+T^{\frac{\alpha}{2}}_{n}\right)},
\end{align}
Note that $\mathrm{\tilde{C}11}$ and $\mathrm{\tilde{C}17}$ follow from Lemma \ref{over_estimator}. Additionally, using C16 implies that $\mathbf{T}$ is non-negative such that $T[n]\geq H$. Therefore, for $H \geq 1$,  $\mathrm{\tilde{C}15}$ is regarded as a convex constraint, and therefore, $(P9)$ being a convex problem can be optimally solved by any known solvers, here, we use CVX \cite{cvx}. Further, for FUJ, the corresponding sub-problem of~\us-trajectory design is similar to  $(P7)$ by replacing \eqref{an_ujt} with $A[n]  = \zeta[n]\gamma_S[n]$, and  following similar approach taken above, the solution of that problem can be obtained.

Finally, for the conventional case WoJ, the sub-problem of~\us~path planning with SWIPT at destination and partially known~\eve~location is reformulated as
\begin{align}\label{osrctrj_wouj}
(P10):~& \stackrel{}{\underset{\mathbf{Q_S}}{\mathrm{maximize}}\sum_{n=1}^{N}\log \Phi_1(\Q_S[n])}\nonumber\\
&~~~~~\text{s.t.}~~~~~ \mathrm{C5-C7, C12},\nonumber\\
&\hspace{-10mm}\mathrm{\widetilde{C}15}:~~ C_n\left(\|\Q_S[n] \hspace{-1mm}-\hspace{-1mm}\W_D\|^2\hspace{-1mm}+\hspace{-1mm}H^2\right)^{-\frac{\alpha}{2}} \hspace{-1mm}+\hspace{-1mm} D_n \geq \Psi_H,
\end{align}
where
\begin{align}\label{abn_wjt}
A_n  &= {\zeta[n]\gamma_S[n]},~~~~~~~~~~~~~B_n = \gamma_S[n],~\nonumber\\
C_n  &= \eta \bar{\beta} \left(1\hspace{-1mm}-\hspace{-1mm}\zeta[n]\right)P_S[n],~~~D_n \hspace{-1mm}=\hspace{-1mm} \eta \left(1\hspace{-1mm}-\hspace{-1mm}\zeta[n]\right)N_0.
\end{align}
which is non-convex because of Lemma \ref{ccv} or non-convex constraint $\mathrm{\widetilde{C}15}$, and therefore we obtain a convex approximated problem of $(P10)$ as
\begin{align}\label{osrctrj_wouj_cvx}
(P11):~& \stackrel{}{\underset{\mathbf{Q_S}, \mathbf{T}, \mathbf{U}}{\mathrm{maximize}}}~\sum_{n=1}^{N}\left[-
\hat{A}_nT[n]\hspace{-1mm}-\hspace{-1mm}\log\left(1\hspace{-1mm}+\hspace{-1mm}B_nU^{-\frac{\alpha}{2}}[n] \right) \right] \nonumber\\
&~~~~\text{s.t.}~~~~~ \mathrm{C5-C7, C12},\nonumber\\
&\hspace{-13mm}\mathrm{\widetilde{C}15}:~\left[\frac{\Psi_H-D_n}{C_n}\right]_+ T^{\frac{\alpha}{2}}[n] \leq 1, \nonumber\\
&\hspace{-13mm}\mathrm{C16}:\|\Q_S[n] -\W_D\|^2+H^2 - T[n] \leq 0,\nonumber\\
&\hspace{-13mm}\mathrm{\tilde{C}17}:2R_E \|\Q_S[n] \hspace{-1mm}-\hspace{-1mm} \hat{\W}_E \|\hspace{-1mm}-\hspace{-1mm}2\hspace{-1mm}\left(\hspace{-0.5mm}{\Q}^k_S[n] \hspace{-1mm}-\hspace{-1mm} {\hat{\W}_E}\hspace{-0.5mm}\right)^\dagger\hspace{-2mm}{\Q_S[n]} \hspace{-1mm}-\hspace{-1mm} U[n]  \hspace{-1mm}+\hspace{-1mm} \widetilde{\mathrm{H}}\hspace{-1mm} \leq \hspace{-1mm}0,
\end{align}
Now $(P11)$ is convex. With an initial point $\left(\mathbf{Q}^k_\mathbf{S}, \mathbf{T}^k, \mathbf{U}^k\right)$, we can solve it optimally with CVX.

\subsection{Optimal UAV-jammer trajectory design}
We are finally after designing an optimal trajectory of~\uj, provided that $\left(\mathbf{P_S, P_J}, \pmb{\zeta}, \mathbf{Q_S}\right)$ are given. For GJT, we formulate the sub-problem of~\uj-trajectory design as
\begin{align}\label{ojmrtrj}
(P12):~&\stackrel{}{\underset{\mathbf{Q_J}}{\mathrm{maximize}}~\sum_{n=1}^{N}\log_2\Phi_2(\Q_J[n])} \nonumber\\
&~~\text{s.t.}~~~~~ \mathrm{C8-C11,~~C13},\nonumber\\
&\hspace{-14mm}\mathrm{\tilde{C}15}:~~E_n+F_n\left(\|\Q_J[n] -\W_D\|^2+H^2\right)^{-\frac{\alpha}{2}} \geq \Psi_H,  
\end{align}
where for $\forall n \in \{1,2,\cdots,N\}$, we have
\begin{align}
\Phi_2(\Q_J[n])\hspace{-1mm}=\hspace{-1mm}\frac{1\hspace{-1mm}+\hspace{-1mm}\frac{A_n}{B_n\left(\|\Q_J[n] -\W_D\|^2+H^2\right)^{-\frac{\alpha}{2}}+1}}{1\hspace{-1mm}+\hspace{-1mm}\frac{C_n}{D_n \left(\|\Q_J[n] - \hat{\W}_E \|^2 +2R_E \|\Q_J[n] - \hat{\W}_E \|+ \tilde{H}^2\right)^{-\frac{\alpha}{2}} +1}},
\end{align}
\begin{align}
\hspace{-5mm}A_n \hspace{-1mm}=\hspace{-1mm} \zeta[n]\gamma_S[n] \left(\|\Q_S[n] \hspace{-1mm}-\hspace{-1mm}\W_D\|^2\hspace{-1mm}+\hspace{-1mm}H^2\right)^{-\frac{\alpha}{2}}\hspace{-2mm},~B_n \hspace{-1mm}=\hspace{-1mm} \zeta[n]\gamma_J[n], \label{an_ujt_ojmrtrj}\\
\hspace{-5mm}C_n \hspace{-1mm}=\hspace{-1mm} \gamma_S[n] \left(\|\Q_S[n] \hspace{-1mm}-\hspace{-1mm} \hat{\W}_E \|^2 \hspace{-1mm}- \hspace{-1mm}2R_E \|\Q_S[n] \hspace{-1mm}-\hspace{-1mm} \hat{\W}_E \|\hspace{-1mm}+\hspace{-1mm} \tilde{H}^2 \hspace{-1mm}\right)^{-\frac{\alpha}{2}}\hspace{-4mm}, \label{cn_ujt_ojmrtrj} \\
\hspace{-5mm}E_n  \hspace{-1mm}= \hspace{-1mm} \eta \left(1 \hspace{-1mm}- \hspace{-1mm}\zeta[n]\right) \hspace{-1mm}\left[ P_S[n]  \bar{\beta}\left(\|\Q_S[n]  \hspace{-1mm}- \hspace{-1mm}\W_D\|^2 \hspace{-1mm}+ \hspace{-1mm}H^2\right)^{-\frac{\alpha}{2}} \hspace{-1mm}+ \hspace{-1mm}N_0\right], \label{en_ujt_ojmrtrj}
\\
\hspace{-5mm}D_n \hspace{-1mm}=\hspace{-1mm}\gamma_J[n],~~~~~~~~F_n = \eta \bar{\beta} \left(1-\zeta[n]\right)P_J[n]. \label{fn_ujt_ojmrtrj}
\end{align}
Reformulating problem $(P12)$ by introducing the slack variables $\mathbf{S} = \{S[1], S[2], \cdots, S[N]\}$ and $\mathbf{V} = \{V[1], V[2], \cdots, V[N]\}$ yields
\begin{align}\label{ojmrtrj_1}
\hspace{-1mm}(P13):&\stackrel{}{\underset{\mathbf{Q_J, S, V}}{\mathrm{maximize}}\sum_{n=1}^{N}
\log \left(\frac{1+\frac{A_n}{B_nS^{-\frac{\alpha}{2}}[n]+1}}{(1+\frac{C_n}{D_nV^{-\frac{\alpha}{2}}[n]+1}} \right)}\nonumber\\
&~~~~~\text{s.t.}~~~~~ \mathrm{C8-C11,~~C13},\nonumber\\
&\hspace{-10mm}\mathrm{\widetilde{C}15}: \left[{\Psi_H-E_n}\right]_+S[n]^{\frac{\alpha}{2}} \leq F_n, \nonumber\\
&\hspace{-10mm}\mathrm{C16}:\|\Q_J[n] \hspace{-1mm}-\hspace{-1mm}\W_D\|^2\hspace{-1mm}+\hspace{-1mm}H^2 \hspace{-1mm}-\hspace{-1mm} S[n] \hspace{-1mm}\geq \hspace{-1mm}0,~~\mathrm{C17}:~S[n]\hspace{-1mm} \geq \hspace{-1mm}0,\nonumber\\
&\hspace{-10mm}\mathrm{C18}:\|\Q_J[n] \hspace{-1mm}-\hspace{-1mm} \hat{\W}_E \|^2 \hspace{-1mm}+\hspace{-1mm}2R_E \|\Q_J[n] \hspace{-1mm}-\hspace{-1mm} \hat{\W}_E \|\hspace{-1mm}+\hspace{-1mm} \tilde{H}^2 \hspace{-1mm}-\hspace{-1mm} V[n] \hspace{-1mm}\leq \hspace{-1mm}0,
\end{align}

\begin{lemma}\label{cvx}
Define the bivariate function $f(x,y) \hspace{-1mm}=\hspace{-1mm} \log\left(1\hspace{-1mm}+\hspace{-1mm}a_1 \exp(x)\right)\hspace{-1mm}+\hspace{-1mm}\log\left(1\hspace{-1mm}+\hspace{-1mm}a_2 \exp(y)\right), x, y>0$ with the non-negative parameters $a_1$ and $a_2$ and the constraint $a \geq 1$. 
By taking the first and second derivative of the function with respect to the variable $x$ and obtaining the corresponding gradient and Hessian of $f$, one can reach at
	\begin{align}
	\nabla(f) = \mathrm{D}f =\left[\frac{a_1e^x}{1+a_1 \exp(x)}, \frac{a_2e^y}{1+a_2 \exp(y)}\right]^\dagger,
	\end{align}
		\begin{align}
		\mathrm{H}(f) = \mathrm{D^2}f = \left[\begin{array}{cc}
		\frac{a_1 e^x}{\left[1+a_1 \exp(x)\right]^2}& 0 \\ 
		0 & \frac{a_1 e^y}{\left[1+a_1 \exp(y)\right]^2}
		\end{array}\right], 
		\end{align}
Since matrix $H$ is positive semidefinite for $t>0$, the function $f(x,y)$ is convex. Therefore, its first Taylor expansion providing a global under-estimator of $f(x,y)$ at point $(x_0,y_0)$ is given by
\begin{align}\label{LB_func}
f(x,y) \geq \nonumber\\
& \hspace{-10mm} f(x_0,y_0) \hspace{-1mm}+\hspace{-1mm} \left[\frac{a_1e^{x_0}}{1\hspace{-1mm}+\hspace{-1mm}a_1 \exp(x_0)}, \frac{a_2e^{y_0}}{1\hspace{-1mm}+\hspace{-1mm}a_2 \exp(y_0)}\right]\hspace{-1mm}(x\hspace{-1mm}-\hspace{-1mm}x_0,y\hspace{-1mm}-\hspace{-1mm}y_0)^\dagger.
\end{align}
\end{lemma}
Based on Lemma \ref{cvx}, the objective function of $(P13)$ is in the form of convex-minus-convex with respect to $\tilde{V}[n] = \frac{\alpha}{2}\log V[n]$ and  $\tilde{S}[n] = \frac{\alpha}{2}\log S[n]$, i.e., it is still non-convex.
Hence, the approximated convex problem corresponding to $(P13)$ can be obtained as
\begin{align}\label{ojmrtrj_cvx_eff}
(P14):&\stackrel{}{\underset{\mathbf{Q_J, \widetilde{S}, \widetilde{V} }}{\mathrm{maximize}}\sum_{n=1}^{N}f_{LB}[n] \hspace{-1mm}-\hspace{-1mm} \log\hspace{-1mm}\left(\hspace{-0.5mm}1\hspace{-1mm}+\hspace{-1mm}b_{1}e^{\tilde{S}[n]}\right)\hspace{-1mm}-\hspace{-1mm}\log\hspace{-1mm}\left(\hspace{-0.5mm}1\hspace{-1mm}+\hspace{-1mm}b_{2}e^{\tilde{V}[n]}\right)}\nonumber\\
&~~~\text{s.t.}~~~~~ \mathrm{C8-C10,~~C13},\nonumber\\
&\hspace{-10mm} \mathrm{\tilde{C}11}:~\tilde{D}^2 \hspace{-1mm}+\hspace{-1mm} \|{\Q}^k_J\|^2 \hspace{-1mm}-\hspace{-1mm} 2\left({\Q}^k_J \hspace{-1mm}-\hspace{-1mm} {\Q}^k_S\right)^\dagger{\Q}_J[n] \hspace{-1mm}-\hspace{-1mm} \|{\Q^k_S}\|^2 \hspace{-1mm} \leq \hspace{-1mm}0,\nonumber\\
&\hspace{-10mm}\mathrm{\widetilde{C}15}:~\left[{\Psi_H\hspace{-1mm}-\hspace{-1mm}E_n}\right]_+\exp(\tilde{S}[n]) \hspace{-1mm}\leq \hspace{-1mm}F_n, \nonumber\\
&\hspace{-10mm}\mathrm{\tilde{C}16}:~2\left({\Q}^k_J[n] \hspace{-1mm}-\hspace{-1mm} {{\W}_D}\right)^\dagger\hspace{-1mm}{\Q_J[n]} \hspace{-1mm}-\hspace{-0.5mm} \exp\left(\hspace{-1mm}\frac{2}{\alpha}\tilde{S}[n]\hspace{-1mm}\right)  +H_1 \hspace{-1mm}\geq\hspace{-1mm} 0,\nonumber\\
&\hspace{-10mm}\mathrm{\tilde{C}18}:~\|\Q_J[n] \hspace{-1mm}-\hspace{-1mm} \hat{\W}_E \|^2 \hspace{-1mm}+\hspace{-1mm}2R_E \|\Q_J[n] \hspace{-1mm}-\hspace{-1mm} \hat{\W}_E \|\hspace{-1mm}+\hspace{-1mm} \tilde{H}^2 \hspace{-1mm}\leq \hspace{-1mm} I_n \hspace{-1mm}+\hspace{-1mm} J_n \tilde{V}[n],
\end{align}
where $H_1 = -\|{\Q^k_J[n]}\|^2+ \|{\hat{\W}_D}\|^2 + {H}^2$ and the concave lower-bound function $f_{LB}$ is given by
\begin{align}
    f_{LB}[n] \treq \frac{a_1\exp\hspace{-0.5mm}\left({\widetilde{S}^k[n]}\hspace{-0.5mm}\right)}{1+a_1 \exp\left({\widetilde{S}^k[n]}\hspace{-0.5mm}\right)}\widetilde{S}[n] \hspace{-1mm}+\hspace{-1mm} \frac{a_2\exp\hspace{-0.5mm}\left({\widetilde{V}^k[n]}\hspace{-0.5mm}\right)}{1+a_2 \exp\hspace{-0.5mm}\left({\widetilde{V}^k[n]}\hspace{-0.5mm}\right)}\widetilde{V}[n],
\end{align}
where $a_1 = \frac{1+A_n}{B_n}, a_2 = \frac{1}{D_n}$, $b_1 = \frac{1}{B_n}, b_2 = \frac{1+C_n}{D_n}$, $I_n=\left(1-\frac{2}{\alpha}\tilde{V}_k[n]\right)\exp\left(\frac{2}{\alpha}\tilde{V}_k[n]\right)$, and $J_n= \frac{2}{\alpha}\exp\left(\frac{2}{\alpha}\tilde{V}_k[n]\right)$. Note that  constraints $\mathrm{\tilde{C}11}$, $\mathrm{\tilde{C}16}$, and $\mathrm{\tilde{C}18}$ are obtained by substituting the non-convex terms of the left hand side constraints C11, C16, and C18 of $(P13)$ with their approximated convex expressions using Lemma \ref{over_estimator}. Since $(P14)$ is now convex, we use CVX and \cite{cvxquad} to solve it, given an initial point $(\mathbf{Q}_\mathbf{J}^k, \widetilde{\mathbf{S}}^k, \widetilde{\mathbf{V}}^k)$, where the superscript $\mathrm{k}$ denotes iteration index.
Further, to optimize~\uj-trajectory for FUJ, we solve $(P13)$ by removing the terms involving $\widetilde{S}[n]$ from its objective function.

{
\subsection{Overall algorithm}
In order to solve problem (P1) by using BCD method for the jammer-included scenarios, we jointly optimize UAVs' transmit power $\mathbf{P}_\mathbf{S}$  and $\mathbf{P}_\mathbf{J}$, destination's PSR factor $\pmb{\zeta}$, as well as UAV-source and UAV-jammer's trajectories $\mathbf{Q}_\mathbf{S}$ and $\mathbf{Q}_\mathbf{J}$ alternatively via solving sub-problems (P2), (P3), (P6), (P7), and (P12), respectively. We summarize the detail of overall iterative solution in Algorithm \ref{algorithm}.}

\begin{algorithm}[t]
\caption{\label{algorithm}{Proposed iterative algorithm}}
			{
			1:~\textbf{Initialize:}~
			Set initial feasible points $\mathbf{P}^{(0)}_\mathbf{S}$, $\mathbf{P}^{(0)}_\mathbf{J}$, $\pmb{\zeta}^{(0)}$, $\mathbf{Q}^{(0)}_\mathbf{S}$, and $\mathbf{Q}^{(0)}_\mathbf{J}$, as well as put the initial values of slack variables   $\mathbf{T}^{(0)}$ and $\mathbf{U}^{(0)}$, $\mathbf{\tilde{S}}^{(0)}$ and $\mathbf{\tilde{V}}^{(0)}$, and let $k=0$;\\
			2:~\textbf{Repeat:}\\
			3:~$k\gets k+1;$\\
			4:~Given $\mathbf{P}^{(k-1)}_\mathbf{J}$, $\pmb{\zeta}^{(k-1)}$, $\mathbf{Q}^{(k-1)}_\mathbf{S}$, and $\mathbf{Q}^{(k-1)}_\mathbf{J}$
		    solve $(\mathrm{P2})$ using \eqref{optimalsrcpow_wj} updating  $\mathbf{P}^{(k)}_\mathbf{S}$;\\
			5:~Given $\mathbf{P}^{(k)}_\mathbf{S}$, $\mathbf{P}^{(k-1)}_\mathbf{J}$, $\pmb{\zeta}^{(k-1)}$, $\mathbf{Q}^{(k-1)}_\mathbf{S}$, and $\mathbf{Q}^{(k-1)}_\mathbf{J}$, solve $(\mathrm{P4})$ via updating $\mathbf{P}^{(k)}_\mathbf{J}$ using \eqref{Pjstar};\\ 
			6:~Given $\mathbf{P}^{(k)}_\mathbf{S}$,  $\mathbf{P}^{(k)}_\mathbf{J}$, $\mathbf{Q}^{(k-1)}_\mathbf{S}$, and $\mathbf{Q}^{(k-1)}_\mathbf{J}$, update  $\pmb{\zeta}^{(k)}$ using \eqref{zeta_opt};\\
			7:~Given  $\mathbf{P}^{(k)}_\mathbf{S}$, $\mathbf{P}^{(k)}_\mathbf{J}$, $\pmb{\zeta}^{(k)}$, $\mathbf{Q}^{(k-1)}_\mathbf{S}$,  $\mathbf{Q}^{(k-1)}_\mathbf{J}$, $\mathbf{T}^{(k-1)}$, and $\mathbf{U}^{(k-1)}$ solve $(\mathrm{P9})$ for GJT/FUJ and 		$(\mathrm{P11})$ for WoJ, updating  $\mathbf{Q}^{(k)}_\mathbf{S}$, $\mathbf{T}^{(k)}$, and $\mathbf{U}^{(k)}$;\\
			8:~Given $\mathbf{P}^{(k)}_\mathbf{S}$, $\mathbf{P}^{(k)}_\mathbf{J}$, $\pmb{\zeta}^{(k)}$, $\mathbf{Q}^{(k)}_\mathbf{S}$,  $\mathbf{Q}^{(k-1)}_\mathbf{J}$, $\mathbf{\tilde{S}}^{(k-1)}$, and $\mathbf{\tilde{V}}^{(k-1)}$ solve  $(\mathrm{P14})$ updating  $\mathbf{Q}^{(k)}_\mathbf{J}$, $\mathbf{\tilde{S}}^{(k)}$, and $\mathbf{\tilde{V}}^{(k)}$;\\			
		    9:~\textbf{Until} the absolute increase of the objective function is below the threshold $\epsilon$;\\
			10:~\textbf{Return:}\\
			$\mathbf{P}^\star_\mathbf{S} \gets \mathbf{P}^{(k)}_\mathbf{S}$,
			$\mathbf{P}^\star_\mathbf{J} \gets \mathbf{P}^{(k)}_\mathbf{J}$,
			$\pmb{\zeta}^\star \gets \pmb{\zeta}^{(k)}$,
			$\mathbf{Q}^\star_\mathbf{S} \gets \mathbf{Q}^{(k)}_\mathbf{S}$,
			$\mathbf{Q}^\star_\mathbf{J} \gets \mathbf{Q}^{(k)}_\mathbf{J}$;}
\end{algorithm}

{
Now, aiming at convergence analysis of Algorithm \ref{algorithm} let define the objective value of original problem; i.e., (P1), at iteration $k$ as $\bar{R}\left(\mathbf{P}^k_\mathbf{S},\mathbf{P}^k_\mathbf{J}, \pmb{\zeta}^k, \mathbf{Q}^k_\mathbf{S}, \mathbf{Q}^k_\mathbf{J}\right)$. 
Similar definitions are taken for the objective values of problems (P4), (P9), and (P14) defined as  $\Theta_{lb}\hspace{-1mm}\left(\mathbf{P}^{k}_\mathbf{S},\mathbf{P}^k_\mathbf{J}, \pmb{\zeta}^k, \mathbf{Q}^k_\mathbf{S}, \mathbf{Q}^k_\mathbf{J}\right)$, 
$\Xi_{lb}\hspace{-1mm}\left(\mathbf{P}^{k}_\mathbf{S},\mathbf{P}^k_\mathbf{J}, \pmb{\zeta}^k, \mathbf{Q}^k_\mathbf{S}, \mathbf{Q}^k_\mathbf{J}\right)$,  $\Omega_{lb}\hspace{-1mm}\left(\mathbf{P}^{k}_\mathbf{S},\mathbf{P}^k_\mathbf{J}, \pmb{\zeta}^k, \mathbf{Q}^k_\mathbf{S}, \mathbf{Q}^k_\mathbf{J}\right)$, respectively. Now, we prove the convergence of Algorithm \ref{algorithm} in what follows. 
\begin{align*}
\bar{R}&\left(\mathbf{P}^k_\mathbf{S},\mathbf{P}^k_\mathbf{J}, \pmb{\zeta}^k, \mathbf{Q}^k_\mathbf{S}, \mathbf{Q}^k_\mathbf{J}\right)
\stackrel{(a)}{\leq}  \bar{R}\left(\mathbf{P}^{(k+1)}_\mathbf{S},\mathbf{P}^k_\mathbf{J}, \pmb{\zeta}^k, \mathbf{Q}^k_\mathbf{S}, \mathbf{Q}^k_\mathbf{J}\right)\nonumber\\
&\stackrel{(b)}{=} \Theta_{lb}\left(\mathbf{P}^{(k+1)}_\mathbf{S},\mathbf{P}^k_\mathbf{J}, \pmb{\zeta}^k, \mathbf{Q}^k_\mathbf{S}, \mathbf{Q}^k_\mathbf{J}\right)\nonumber\\
&\stackrel{(c)}{\leq} \Theta_{lb}\left(\mathbf{P}^{(k+1)}_\mathbf{S},\mathbf{P}^{(k+1)}_\mathbf{J}, \pmb{\zeta}^k, \mathbf{Q}^k_\mathbf{S}, \mathbf{Q}^k_\mathbf{J}\right)\nonumber\\
&\stackrel{(d)}{\leq} \bar{R}\left(\mathbf{P}^{(k+1)}_\mathbf{S},\mathbf{P}^{(k+1)}_\mathbf{J}, \pmb{\zeta}^{k}, \mathbf{Q}^k_\mathbf{S}, \mathbf{Q}^k_\mathbf{J}\right)\nonumber\\
&\stackrel{(e)}{\leq} \bar{R}\left(\mathbf{P}^{(k+1)}_\mathbf{S},\mathbf{P}^{(k+1)}_\mathbf{J}, \pmb{\zeta}^{(k+1)}, \mathbf{Q}^k_\mathbf{S}, \mathbf{Q}^k_\mathbf{J}\right)\nonumber\\
&\stackrel{(f)}{=} \Xi_{lb}\left(\mathbf{P}^{(k+1)}_\mathbf{S},\mathbf{P}^{(k+1)}_\mathbf{J}, \pmb{\zeta}^{(k+1)}, \mathbf{Q}^k_\mathbf{S}, \mathbf{Q}^k_\mathbf{J}\right)\nonumber\\
\end{align*}
\begin{align}\label{convergence}
&\stackrel{(g)}{\leq} \Xi_{lb}\left(\mathbf{P}^{(k+1)}_\mathbf{S},\mathbf{P}^{(k+1)}_\mathbf{J}, \pmb{\zeta}^{(k+1)}, \mathbf{Q}^{(k+1)}_\mathbf{S}, \mathbf{Q}^k_\mathbf{J}\right)\nonumber\\
&\stackrel{(h)}{\leq} \bar{R}\left(\mathbf{P}^{(k+1)}_\mathbf{S},\mathbf{P}^{(k+1)}_\mathbf{J}, \pmb{\zeta}^{(k+1)}, \mathbf{Q}^{(k+1)}_\mathbf{S}, \mathbf{Q}^k_\mathbf{J}\right)\nonumber\\
&\stackrel{(i)}{=} \Omega_{lb}\left(\mathbf{P}^{(k+1)}_\mathbf{S},\mathbf{P}^{(k+1)}_\mathbf{J}, \pmb{\zeta}^{(k+1)}, \mathbf{Q}^{(k+1)}_\mathbf{S}, \mathbf{Q}^k_\mathbf{J}\right)\nonumber\\
&\stackrel{(j)}{\leq} \Omega_{lb}\left(\mathbf{P}^{(k+1)}_\mathbf{S},\mathbf{P}^{(k+1)}_\mathbf{J}, \pmb{\zeta}^{(k+1)}, \mathbf{Q}^{(k+1)}_\mathbf{S}, \mathbf{Q}^{(k+1)}_\mathbf{J}\right)\nonumber\\
&\stackrel{(k)}{\leq} \bar{R}\left(\mathbf{P}^{(k+1)}_\mathbf{S},\mathbf{P}^{(k+1)}_\mathbf{J}, \pmb{\zeta}^{(k+1)}, \mathbf{Q}^{(k+1)}_\mathbf{S}, \mathbf{Q}^{(k+1)}_\mathbf{J}\right),
\end{align}
where the inequalities $(a)$, $(c)$, $(e)$, $(g)$, and $(j)$ all follow from the definition of the optimal solution to the problems (P2), (P4), (P6), (P9), and (P14), respectively. Besides, the equality $(b)$ holds since the first order Taylor approximation is adopted and that the objective function of problems (P3) and (P4) share the same value at $\mathbf{P}^k_\mathbf{J}$. Similar justifications can be explained for the equalities $(f)$ and $(i)$ at points $\mathbf{Q}^{k}_\mathbf{S}$ and $\mathbf{Q}^{k}_\mathbf{J}$, respectively. Further, $(d)$, $h$, and $(k)$ follow from the fact that the objective functions of problems (P4), (P9), and (P14) are tight lower-bounds to that of (P3), (P7), and (P12), respectively. The last inequality in \eqref{convergence} indicates that the objective value of (P1) is non-decreasing over the  iteration index. As well as that, the optimal value of (P1) is finite, i.e., the optimal ASR is upper bounded by a finite value, which means the proposed iterative Algorithm \ref{algorithm} is guaranteed to converge.  Due to the convexity of the approximated sub-problems (P4), (P9), and (P14), the proposed algorithm is appropriate for UAV applications as it can be efficiently implemented in practice as having a complexity of  $\mathcal{O}(kN^{m})$, where $m$ is the number of variable blocks, which means the solution can be obtained at worst-case in polynomial time.}
\section{Numerical Results}\label{Sec:Simu}
In simulations, unless otherwise stated, we adopt the following parameters. {The mission time duration, in consistency with \cite{Fuhui2018Com} is chosen $T=2s$ which is discretized into $N=100$ equal time slots to balance the accuracy and computational complexity}, the total power budget divided equally between UAVs is $ P^{tot}_S+P^{tot}_J = 20$ dBm with the maximum instantaneous transmit power of $2$mW and PAPR ratio of $4$, leading to average transmission power $0.5$mW of each. Considering the normalized transmission bandwidth, we set $N_0 = -40$ dBm, $\Psi_H = -20$ dBm with power conversion efficiency factor $\eta = 0.7$ in (\ref{eh}). We set $\gamma_0 = 40$dB and path-loss exponent $\alpha = 2.5$. We set $H=1.5$ with $R = 2.5H$ (radius of permitted flying circular  region centered at~\des), $\W_D=(0,0)$, $L=\frac{R}{2}$ (distance between ground destination and geometric center of the eavesdropper), $\hat{\W}_E = (L,0)$, where the exact location of~\eve~is a random point within the circular region centered at $\hat{\W}_E$ with radius $R_E = \frac{H}{5}$, and safety distance between UAVs $\tilde{D} = \frac{H}{10}$. 
$\bar{\beta}$ is obtained by averaging over $10^5$ channel realizations over the area of interest.
In all plots, we compare FUJ, GJT, and WoJ in terms of the following aspects:
\begin{itemize}
\item {convergence of the proposed iterative algorithm, demonstrated by variation of average secrecy rate with respect to iteration index wherein we utilized absolute error function $f_{err}(k)$= $\|{{{\bar{R}}_{sec}^{opt, k}}} - {{{\bar{R}}_{sec}^{opt, k-1}}}\|$
as the termination criteria similar to \cite{sun2019Swipt}},
\item optimal UAVs' trajectory,
\item instantaneous secrecy rate,
\item ASR and average harvested energy (AHE) at~\des,
\item {\em instantaneous secrecy energy efficiency} (ISEE) defined as the ratio between $\bar{R}_{sec}^{opt}[n]$ and ${P_S[n]+P_J[n]}$,
\item UAVs' transmit power over flying horizon,
\item impact of estimated location of~\eve~on the ASR and AHE 
\item {\em harvested power efficiency} defined as $\frac{\tilde{P}_H[n]}{{P_S[n]+P_J[n]}}$.
\end{itemize}
In particular, in optimal trajectory comparisons, we adopt the so-called {\em baseline scheme} for UAVs initial trajectory; i.e., both~\us~and~\uj~fly with their maximum speeds towards as close as~\des~and~the geometric center of estimated location of~\eve, respectively. Then, both UAVs hover above the corresponding points as long as possible in order to send the data and  conduct jamming transmission, respectively, followed by heading with their maximum speeds towards final location, provided that the mission time is sufficient. Otherwise, they turn from a midway heading towards the final locations.

\begin{figure}[t]
	\center{\includegraphics[ width=0.85\columnwidth ]{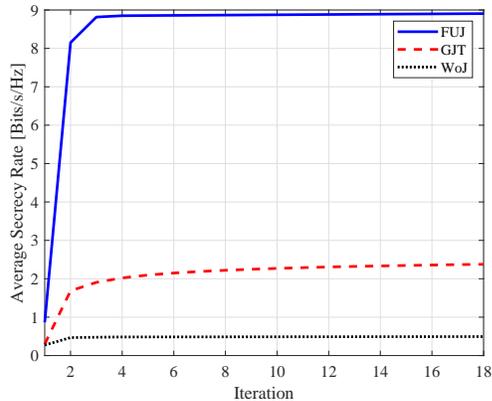}}
	\caption{\label{sim1} Average secrecy rate against iteration.} 
\end{figure}
{Fig. \ref{sim1} illustrates the convergence plot of the proposed iterative algorithms for FUJ, GJT, and WoJ. We plot the ASR as the number of iteration $k$ varies. We see all schemes converge with terminating threshold $\epsilon = 10^{-2}$, validating our analysis in terms of convexity of the approximated sub-problems. It should be mentioned that  Algorithm \ref{algorithm} for all the scenarios converges quite quickly in few iterations making it an efficient solution for the considered UAV application.}


\begin{figure}[t]
	\center{\includegraphics[ width=0.85\columnwidth ]{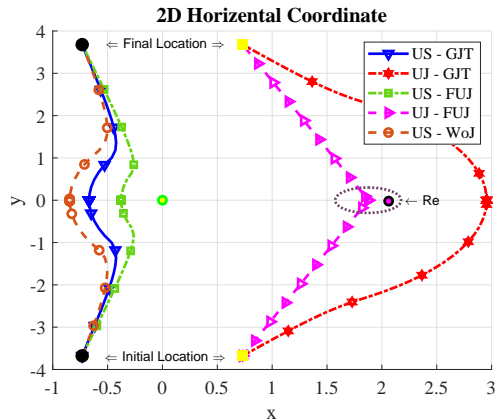}}
	\caption{\label{sim2} Optimal trajectory of UAVs for GJT, FUJ, and WoJ scenarios.}             
\end{figure}


Fig. \ref{sim2} illustrates the optimal UAVs' trajectory for FUJ, GJT, and WoJ using the proposed sequential algorithm. Note the {\em green-edge} and {\em black-edge} circles denote the exact location of~\des~and~\eve, respectively. We observe that, for FUJ scheme,~\us~gets the closest to~\des~among all, with substantially improved ASR. For FUJ, the operation time and energy constraints can make~\uj~head directly to the best possible position for jamming, which is much shorter than GJT.




	

\begin{figure}[t]
	\center{\includegraphics[ width=0.85\columnwidth ]{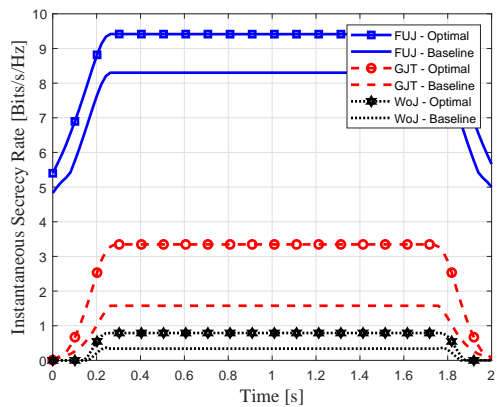}}
	\caption{\label{sim4} Instantaneous secrecy rate verses time}  
\end{figure}
Fig. \ref{sim4} compares ISR of FUJ, GJT, and WoJ using the proposed optimization methods and the {\em aforementioned baseline scheme}, and demonstrates our method leads to a significant performance improvement. We also observe that FUJ brings always positive secrecy rate; nonetheless, WoJ provides zero ISR at the beginning and end of the mission.
{Note that since our objective function formulated to be optimized was the average secrecy rate over the mission time, so the ISR performance is not necessarily expected to be improved at all the mission time, though, we observe significant out-performance  compared to the base-line curves on the whole. Particularly, it can be seen from the optimal curves belong to the FUJ, GJT, and WoJ schemes in Fig. \ref{sim4} that by jointly optimizing transmit power of UAVs as well as their trajectories alongside with the PSR factor we could obtain approximately 2, 1, and 0.5 bits/S/Hz ISR improvements during middle of the mission, respectively. }


\begin{figure}[t]
	\center{\includegraphics[ width=0.85\columnwidth ]{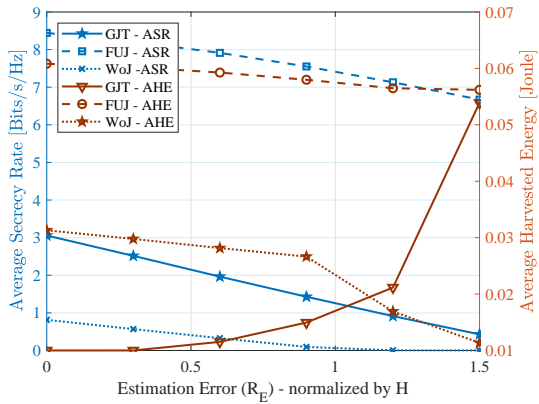}}
	\caption{\label{sim5} Secrecy and energy harvesting  performance against estimation error ($\protect\mathrm{R_E}$).}
\end{figure}

Fig. \ref{sim5} illustrates ASR and AHE at~\des~vs $R_E$ (estimation error of~\eve's location) in FUJ, GJT, WoJ, and demonstrates the resultant ASRs decrease as $R_E$ increases. We observe AHE of WoJ decreases, AHE of FUJ remains approximately unchanged, AHE of GJT  increases. 
This can be interpreted that, as the uncertainty of eavesdropper's location  increases (corresponding to a larger $R_E$), for FUJ and WoJ, UAV~\us~ flies further and has longer distance to~\des, resulting in decreased main link capacity and AHE. However, for GJT, since UAV~\uj~has quite less impact on secrecy as the wiretap link might be better than the main link due to estimation erroneous, UAV~\us~tries to get as close as possible to~\des~in a straight way for improving ASR which, of course, makes AHE increased.


\begin{figure}[t]
	\center{\includegraphics[ width=0.85\columnwidth ]{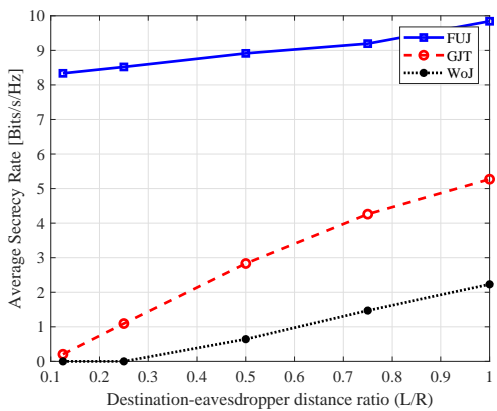}}
	\caption{\label{sim5-1} Average Secrecy Rate vs. destination-eavesdropper distance. y-coordinate of the eavesdropper is set to zero.}
\end{figure}
{Fig. \ref{sim5-1} is presented to draw insight into the impact of the location of the eavesdropper. As it can be clearly seen from the figure, the farther the eavesdropper's location from the destination becomes, the higher the ASR performance is obtained, as expected,  for all the scenarios. Notably, having the highest slop the curve belong to the GJT scheme is more sensitive to this parameter in comparison with the others, which means eavesdropper's location has more impact on the ASR performance of the GJT which should be considered in system design.  Further, when~\eve~gets closer to~\des~the proposed jamming-included scenarios could obtain positive secrecy rate though the WoJ scheme lacks. Particularly, the FUJ scheme regardless of the eve's location provides the best secrecy performance.}


\begin{figure}[t]
	\center{\includegraphics[ width=0.85\columnwidth ]{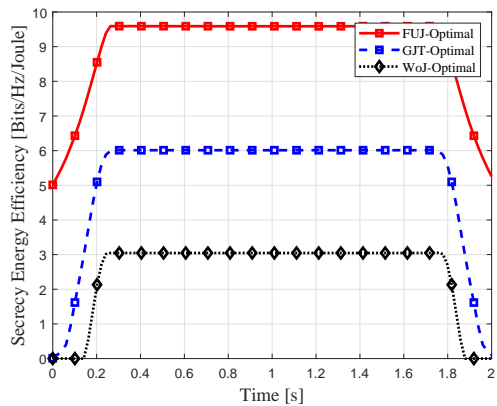}}
	\caption{\label{sim6} Secrecy energy efficiency vs. mission time.}  
\end{figure}
Fig. \ref{sim6} shows ISEE vs. mission time for FUJ, GJT, and WoJ and demonstrates the significant performance improvement of FUJ. This ISEE plot provides a trade-off between ASR and the cost of energy level for communications.  We observe, for all cases, decreasing the distances between (\us,~\uj) and intended ground nodes (\des,~\eve) leads to higher ISEE.

\begin{figure}[t]
	\center{\includegraphics[ width=0.85\columnwidth ]{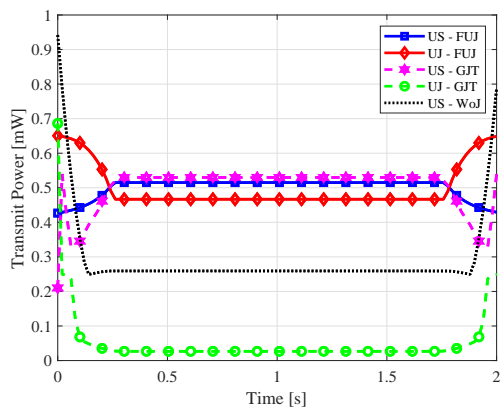}}
	\caption{\label{sim7} {Transmit power vs. time.}}   
\end{figure}
Fig. \ref{sim7} shows UAVs' transmit power over time horizon. For FUJ, at the beginning~\us~decreases its power to against information leakage while~\uj~increases power to satisfy the required minimum energy constraint at destination. When~\us~and~\uj~fly to proper positions for data transmission and jamming, transmit power varies accordingly. For GJT, jamming power remains lowest to avoid degradation of ASR. Finally, WoJ keeps its power resource for the best use when having a better main channel quality with keeping~\us~trajectory to be as far as possible from the estimated location of~\eve. {Interestingly, we observe that even with a significantly lower transmission power of UAV-jammer for the GJT  compared to the UAV-source, the secrecy performance of the jammer-included scenarios could be enhanced.} 


\begin{figure}[t]
	\center{\includegraphics[ width=0.85\columnwidth ]{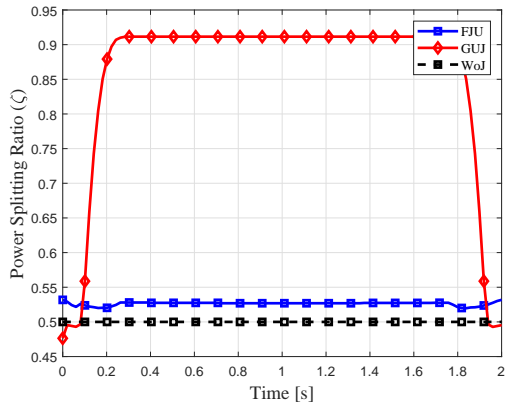}}
	\caption{\label{sim7-1} {Power splitting ratio vs. time.}}   
\end{figure}
Fig. \ref{sim7-1} is provided to demonstrate how the PSR factor varies to make the adequate energy to be harvested by the EH component of the destination for all the three scenarios. We observe that for the GJT the more fraction of the received signals should be dedicated for energy scavenging to satisfy the energy requirement of the destination node  over the time horizon.

\begin{figure}[t]
	\center{\includegraphics[ width=0.85\columnwidth ]{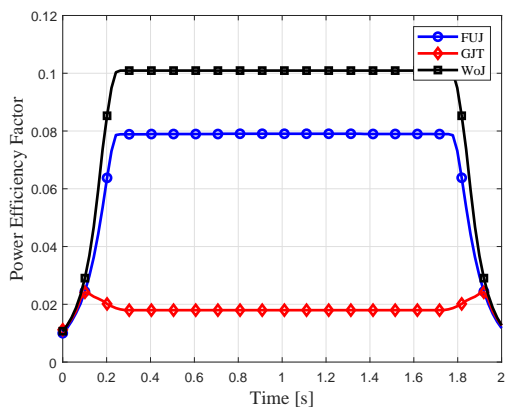}}
	\caption{\label{sim8} Harvested power efficiency}
\end{figure}
Fig. \ref{sim8} illustrates instantaneous harvested power efficiency for FUJ, GJT, and WoJ, with respective fraction of total power budget $P_S[n]+P_J[n]$ and the ratio of total harvested power to the transmit network power obtained as  6.8$\%$, 1.8$\%$ and 7.8$\%$, respectively. We see that, for all cases, energy harvesting constraint is satisfied and the harvested power is well above the minimum requirement $\Psi_H$, particularly WoJ. This indicates that how we can design secure as well as energy efficient UAV-based communications protocols which is a good direction for our future work.
\section{Conclusion}\label{Sec: Conl}
We have considered a 2-UAV based wireless communication system. It consists of two flying cooperative UAVs, a ground destination node  equipped with SWIPT technique, and a passive ground eavesdropper. One UAV acts as source transmitting confidential information to destination, while the other UAV propagates jamming to assist destination with anti-eavesdropping and energy harvesting. Assuming that UAVs have imperfect channel estimation eavesdropper, we have proposed two transmission schemes: FUJ and GJT, transmitting jamming signals that are a priori known and unknown at destination, respectively. Under such setting, we have formulated an average secrecy rate
(ASR) maximization problem in terms of  trajectory design and power controlling, and proposed an iterative algorithm based on the block  coordinated descent and successive convex  approximation. Via this algorithm, we have found the best transmit power and trajectory of both UAVs, as well as the best power splitting ratio of destination. Finally, we have evaluated the proposed schemes by simulations in terms of ASR, ISR, AHE, and demonstrated their effectiveness. In particular, FUJ provides by far the highest ASR improvement compared to GJT and WoJ (the benchmark schemes). 

\appendices
\numberwithin{equation}{section}
\makeatletter 
\newcommand{\section@cntformat}{Appendix \thesection:\ }
\makeatother

\section{Appendix A: Derivation of maximum \texorpdfstring{$I_E$}{} }\label{Appendix A}
First let mention a useful lemma below.
\begin{lemma}\label{Delta_fxy}
	Let define the bivariate function as
	\begin{align}
	f (x,y) = \log\left(1+\frac{\left(x^2 + h\right)^{-a}}{\left(y^2 + h\right)^{-a}+b}\right),
	\end{align}
	where $x,y \geq 0$ and  $a$, $b$, and $h$ are positive constants.  Its gradient can be calculated as
	\begin{equation}\label{grad_fxy}
	\nabla_{x,y}f(x,y) = 
	\begin{bmatrix} 
	-\frac{a}{2 c_1\left(c_1+c_2+n\right)}\\
	\frac{a c_1}{2c_2 \left(c_2+n\right) \left(c_1+c_2+n\right)} 
	\end{bmatrix},
	\end{equation}
	where auxiliary variables defined as $c_1 \treq (H+x)^{-a/2}$ and $c_2 \treq (H+y)^{-a/2}$ are always positive.
	From \eqref{grad_fxy} it follows that the inequalities $f(x,y) > f(x+\epsilon,y)$ and $f(x,y) < f(x,y+\epsilon)$ hold  for any positive-valued $\epsilon$. 
\end{lemma}
Following from Lemma \ref{Delta_fxy} we conclude that the expression given by \eqref{IE}  is a monotonically decreasing function  with respect to the term $\|\Q_S(t) - \W_E \|$ and a monotonically increasing function  with respect to the term $\|\Q_J(t) - \W_E \|$. Then, from linear algebra and applying  the regular and the reverse triangular inequality, one can obtain as {
\begin{align}\label{ub}
\|\Q_J(t) - \W_E \| &\leq  \|\Q_J(t) - \hat{\W}_E \| + \|\hat{\W}_E - \W_E \| \nonumber\\&\leq  \|\Q_J(t) - \hat{\W}_E \| + R_E. 
\end{align}}
and
\begin{align}\label{lb}
\|\Q_S(t) - \W_E \| &\geq   |\|\Q_S(t) - \hat{\W}_E \| - \|\hat{\W}_E - \W_E \||  \nonumber\\
&\geq   |\|\Q_S(t) - \hat{\W}_E \| - R_E|. 
\end{align}     
Then, plugging the lower and upper deterministic expressions respectively  given in \eqref{ub} and \eqref{lb} into \eqref{IE}, leads to the final expression of ${I}^{max}_{E}(t)$ as given in \eqref{IE_max}.
\vspace{-3mm}
\section{Appendix B: proof of Lemma \ref{over_estimator}}\label{Appendix B}
We commence from the concavity of the function $f(x) = -\|\mathrm{x}-\mathrm{a}\|^2$ with the gradient equal to $\nabla f(x) = -2 (\mathrm{x}-\mathrm{a})$ as
\begin{align}
f(x)  &\stackrel{(a)}{\leq} -\|\mathrm{x}^k-\mathrm{a}\|^2 -2 (\mathrm{x}^k-\mathrm{a})(\mathrm{x}-\mathrm{x}^k)\nonumber\\
&=-\|\mathrm{x}^k\|^2 +2\mathrm{a}^\dagger\mathrm{x}^k-\|\mathrm{a}\|^2-2 (\mathrm{x}^k-\mathrm{a})^\dagger(\mathrm{x}-\mathrm{x}^k)\nonumber\\
&=\|\mathrm{x}^k\|^2 -\|\mathrm{a}\|^2-2 (\mathrm{x}^k-\mathrm{a})^\dagger\mathrm{x}
\end{align}
where $(a)$ follows after the fact that the first order Taylor approximation of a concave function is a global affine over-estimator of the function $f(\mathbf{x})$ at the point $\mathrm{x_0}$.


\bibliographystyle{IEEEtran}
\bibliography{MyReferences_IEEE}

\end{document}